\documentclass[
 reprint,
 amsmath,amssymb,
 aps,
 prb,
 longbibliography,
 superscriptaddress,
 floatfix
]{revtex4-2}
\usepackage{graphicx} 
\usepackage{dcolumn}
\usepackage{bm}
\usepackage{physics}
\usepackage[colorlinks=true,allcolors=blue]{hyperref}
\usepackage{multirow}
\usepackage[normalem]{ulem}
\usepackage[dvipsnames]{xcolor}

\begin{document}

\title{Spin current generation via magnetic skyrmion, bimeron, and meron crystals}
\author{Aoi Kajihara}
 \email{a.kajihara@aion.t.u-tokyo.ac.jp}
 \affiliation{Department of Applied Physics, The University of Tokyo, Bunkyo, Tokyo 113-8656, Japan}
 
\author{Shun Okumura}%
 \affiliation{Quantum-Phase Electronics Center (QPEC), The University of Tokyo, Bunkyo, Tokyo 113-8656, Japan}
 \affiliation{RIKEN Center for Emergent Matter Science (CEMS), Wako, Saitama 351-0198, Japan}
 
\author{Yukitoshi Motome}
 \email{motome@ap.t.u-tokyo.ac.jp}
 \affiliation{Department of Applied Physics, The University of Tokyo, Bunkyo, Tokyo 113-8656, Japan}
 
\date{\today}

\begin{abstract}
 Spin current offers a promising route toward energy-efficient and high-speed information processing. Developing efficient methods for their generation remains a central challenge in spintronics. Here, we investigate spin current generation via two-dimensional topological spin textures: a skyrmion crystal (SkX) with out-of-plane magnetization, a bimeron crystal (BmX) with in-plane magnetization, and a meron crystal (MX) with zero net magnetization. We show that these distinct spin textures generate spin currents with characteristic spin polarization directions. In the absence of spin--orbit coupling, the SkX and BmX generate spin currents polarized along their magnetization directions, whereas the MX yields no spin current. Upon introducing spin--orbit coupling, while the behavior of the SkX does not qualitatively change, the BmX generates nonzero spin currents in multiple polarization directions. Notably, the MX, despite its zero net magnetization, exhibits a pronounced spin current with out-of-plane spin polarization, driven by an enhanced spin Berry curvature associated with characteristic band degeneracy. We further demonstrate that the electronic and spin transport properties of each texture are governed by their magnetic symmetries.  Our results highlight the topological spin textures as efficient sources of spin current even without net magnetization, expanding the design for spintronics devices based on topological magnetic metals.
\end{abstract}

\maketitle
\section{Introduction}
A central challenge in spintronics, which aims to manipulate both the spin and charge degrees of freedom of electrons, is how to generate and control spin currents efficiently. Spin currents enable magnetization switching via spin-transfer torque \cite{STT3,STT4,STT5,STTRev}, offering faster memory writing with lower energy consumption compared to magnetic field-based approaches \cite{STT_MRAM1,STT_MRAM2,STT_MRAM3}. This makes spin currents highly promising information carriers. To date, various methods for spin current generation have been developed. For nonmagnetic systems, representative mechanisms include the spin Hall effects in heavy metals or semiconductors with strong spin--orbit coupling (SOC) \cite{SHE1,SHE2,Sinova_Conv,SHE3,SHE4,SHE_Review} and chirality-induced spin selectivity in chiral systems \cite{CISS1,CISS2,CISS3,CISS4,CISS5,CISS_Rev}. Ferromagnetic systems have also been widely studied, where spin currents can be produced not only through electrically driven spin-polarized charge currents \cite{Mott1,Mott2,TwoCurrmodel1,TwoCurrmodel2,SpolC1,SpolC2,STT3,STT4,STT5}, but also via dynamical spin excitations, as exemplified by spin pumping \cite{Spump1,Spump2,Spump3,Spump4,Spump5,Spump6} and the spin Seebeck effect \cite{SSE1,SSE2,SSE3,SSE4,SSE5,SSE6}. More recently, to further enhance the efficiency, unconventional magnets, such as noncollinear antiferromagnets \cite{Noncol1,Noncol2,Noncol3,Noncol4,Noncol5,Noncol6} and altermagnets \cite{almag1,almag2,almag3,almag4,almag5}, have attracted increasing attention.

Among such unconventional magnetic systems, topological spin textures provide a promising platform for spin current generation. These magnetic textures are characterized by nontrivial topological numbers defined by their real-space spin configurations. They comprise a wide variety of types and have been extensively studied both experimentally and theoretically, including one-dimensional chiral solitons \cite{CSs1,CSs2,CSs3,CSs4}, two-dimensional magnetic skyrmions \cite{Sks1,Sks2,Sks3}, and three-dimensional magnetic hedgehogs \cite{HGs1,HGs2} and hopfions \cite{Hopf1,Hopf2,Hopf3}. A notable feature of topological spin textures with noncollinear or noncoplanar spin configurations is their ability to induce emergent electromagnetic fields \cite{HGs1,EMF1,EMF2,EMF3}. These are fictitious electromagnetic fields acting on itinerant electrons coupled to the spin texture. In noncoplanar spin configurations, emergent magnetic fields arise from the Berry phase \cite{Berry} associated with nonzero scalar spin chirality, giving rise to a variety of unusual quantum transport phenomena, such as the topological Hall effect \cite{THE1,THE2,THE3,THE4} and the topological Nernst effect \cite{TNE1,TNE2,TNE3}. By contrast, emergent electric fields are induced by the dynamics of spin textures, leading to characteristic responses exemplified by the spin-motive force \cite{SMF1,SMF2,SMF3,SMF4} and the emergent inductance \cite{EI1,EI2}. Another important characteristic of topological spin textures is the rich variety of magnetic phases, which can be tuned by external fields and temperature \cite{Sks3,Variety1,Variety2,Variety3}. This tunability enables flexible control over distinct magnetic states, offering promising avenues for devices with switchable functionalities \cite{Variety3}. Moreover, in the vicinity of such phase boundaries, competition between different magnetic phases and associated fluctuations can lead to giant responses \cite{Giant1,Giant2,Giant3,Giant4,Giant5}. Owing to these unique features, topological spin textures are highly attractive for spintronics applications, particularly as potentially efficient generators of spin currents.

Indeed, spin current generation using topological spin textures has been examined theoretically for several systems. A representative example is the topological spin Hall effect in magnetic skyrmions \cite{TSHE_sk1,TSHE_sk2,TOHE}, which arises from a topological Hall current that acquires spin polarization due to the net magnetization of the spin texture. Accordingly, a magnetic hopfion, a three-dimensional topological soliton, does not exhibit the topological spin Hall effect since the net emergent magnetic field vanishes \cite{OHE_hop}. By contrast, a transverse pure spin current can be generated in antiferromagnetic textures, such as antiferromagnetic skyrmions \cite{TSHE_AFsk1, TSHE_AFsk2,TOHE,TSHE_AFmr} and antiferromagnetic merons \cite{TSHE_AFmr}; nevertheless, the magnetization and emergent magnetic field cancel globally, while both remain locally nonzero. Despite the strong sensitivity to the underlying spin structure, many aspects of spin current generation remain far from being systematically understood across different types of topological spin textures. Notably, most previous studies have neglected the effect of SOC and have primarily focused on spin currents with out-of-plane spin polarization. These situations motivate us to explore spin current generation across a broader range of topological spin textures, explicitly incorporating SOC and examining spin currents with in-plane polarization components.

In this paper, we theoretically investigate spin current generation arising from the coupling between itinerant electrons and two-dimensional topological spin textures. We consider three types of textures that share a common out-of-plane emergent magnetic field but have distinct magnetization: a skyrmion crystal (SkX) with out-of-plane magnetization, a bimeron crystal (BmX) with in-plane magnetization, and a meron crystal (MX) with zero net magnetization, as shown in Figs.~\hyperlink{fig:spin_texture_bands}{\ref{fig:spin_texture_bands}(a)}, \hyperlink{fig:spin_texture_bands}{\ref{fig:spin_texture_bands}(b)}, and \hyperlink{fig:spin_texture_bands}{\ref{fig:spin_texture_bands}(c)}, respectively. By applying linear response theory to the calculated electronic states, we evaluate the spin conductivities in both longitudinal and transverse channels for all spin-polarization components. In this analysis, we consider both intrinsic contributions independent of scattering mechanisms and dissipative contributions arising from Fermi surfaces. As a result, we find that, in the absence of SOC, the SkX and BmX can generate spin currents polarized along their respective magnetization directions, whereas the MX generates no spin current. Upon introducing Rashba-type SOC, although the behavior for the SkX does not change qualitatively, the BmX generates spin currents with multiple polarization components, and notably, the MX, despite its zero net magnetization, exhibits a pronounced spin current with out-of-plane spin polarization at a particular electron filling due to the enhanced spin Berry curvature arising from specific band degeneracy. In addition, through symmetry analysis based on the spin space group \cite{SSG1,SSG2,SSG3,SSG4,SSG5,SSG6,SSG7} and the magnetic space group, we demonstrate that these results follow the underlying symmetry requirements. Our findings show that topological spin textures can serve as efficient spin current generators even in the absence of net magnetization, thereby expanding the potential of topological magnetic metals for spintronics applications.

The organization of this paper is as follows. In Sec.~\ref{sec:Model}, we introduce the model Hamiltonian and the three topological spin textures considered in this study. Section~\ref{sec:Method} describes the methods used to obtain the electronic structure and transport properties based on linear response theory. In Sec.~\ref{subsec: Bands}, we discuss the electronic band structures and spin splittings for each spin texture. Section~\ref{subsec: transport} provides the charge and spin conductivities, both with and without SOC. In Sec.~\ref{subsec: symmetry}, we examine the consistency of the numerical results with the magnetic symmetries. Finally, in Sec.~\ref{sec: Summary}, we summarize our results and discuss future perspectives.

\section{Model}
\label{sec:Model}
\subsection{Model Hamiltonian}
 We consider a two-dimensional model in which itinerant electrons are coupled to a topological spin texture formed by localized moments on a square lattice. The Hamiltonian is given by
 \begin{equation}
 \hypertarget{eq:Hamiltonian}{}
 \hat{\mathcal{H}}=\hat{\mathcal{H}}_{\mathrm{K}}+\hat{\mathcal{H}}_{\mathrm{SOC}},
 \label{eq:Hamiltonian}
 \end{equation}
 where $\hat{\mathcal{H}}_{\mathrm{K}}$ denotes the kinetic motion of itinerant electrons and their coupling to localized moments, and $\hat{\mathcal{H}}_{\mathrm{SOC}}$ represents the SOC acting on itinerant electrons.

 The first term of Eq.~\hyperlink{eq:Hamiltonian}{(\ref{eq:Hamiltonian})}, which is called the Kondo lattice model or the $s$--$d$ model \cite{double-ex}, is defined as 
 \begin{equation}
    \hypertarget{eq:Kondo}{}
    \hat{\mathcal{H}}_{\mathrm{K}}=-t\sum_{\ev{i,j},\alpha} \left(\hat{c}_{i\alpha}^{\dag}\hat{c}_{j\alpha}+\mathrm{h.c.} \right)-J\sum_{i,\alpha,\beta} \mathbf{S}_i\cdot \hat{c}_{i\alpha}^{\dag}\bm{\sigma}_{\alpha\beta}\hat{c}_{i\beta},
    \label{eq:Kondo}
 \end{equation}
 where $\hat{c}_{i\alpha}^{\dag}\ (\hat{c}_{i\alpha})$ is the creation (annihilation) operator of an electron with spin index $\alpha$ at lattice site $i$. The first term of Eq.~\hyperlink{eq:Kondo}{(\ref{eq:Kondo})} describes electron hopping between all nearest-neighbor sites $\ev{i,j}$ with amplitude $t$. The second term represents the $s$--$d$ coupling between localized spins and itinerant electrons with the strength of $J$. Here, the localized spin at site $i$ is denoted by $\mathbf{S}_i$, and the vector $\bm{\sigma}=(\sigma^x,\sigma^y,\sigma^z)$ represents the vector of Pauli matrices for the spin degrees of freedom of itinerant electrons. In this study, the localized spins are treated as classical vectors with unit length, $\abs{\mathbf{S}_i} = 1$.  
 
  For the SOC in the second term of Eq.~\hyperlink{eq:Hamiltonian}{(\ref{eq:Hamiltonian})}, we adopt the Rashba type, a representative mechanism influencing spin transport in two-dimensional systems, as exemplified by the universal intrinsic spin Hall effect \cite{Sinova_Conv}. The Rashba-type SOC Hamiltonian is defined as
\begin{equation}
    \hat{\mathcal{H}}_{\mathrm{SOC}}=\mathrm{i} \lambda_{\mathrm{R}}\sum_{\ev{i,j},\alpha,\beta}\hat{c}_{i\alpha}^{\dag}\left[\mathbf{\hat{z}}\cdot(\mathbf{d}_{ij}\times \bm{\sigma}_{\alpha\beta})\right]\hat{c}_{j\beta}+\mathrm{h.c.},
\end{equation}
where $\lambda_{\mathrm{R}}$ is the Rashba parameter, $\mathbf{d}_{ij}=\mathbf{r}_i-\mathbf{r}_j$ is the vector from site $j$ to $i$ ($\mathbf{r}_i$ denotes the position vector of site $i$), and $\mathbf{\hat{z}}$ is the unit vector perpendicular to the $xy$ plane on which the two-dimensional lattice is defined. This term arises from an out-of-plane potential gradient that breaks mirror symmetry of the crystal structure with respect to the $xy$ plane. 

In the following calculations, the lattice constant is set to unity. The $s$--$d$ coupling strength is fixed at $J=5t$, representing a strong-coupling regime where the spins of itinerant electrons are almost aligned with the localized moments. For calculations with SOC, the Rashba parameter is set to $\lambda_{\mathrm{R}}=0.1t$.

 \begin{figure*}[htbp]
    \centering
    \hypertarget{fig:spin_texture_bands}{}
    \includegraphics[width=\textwidth]{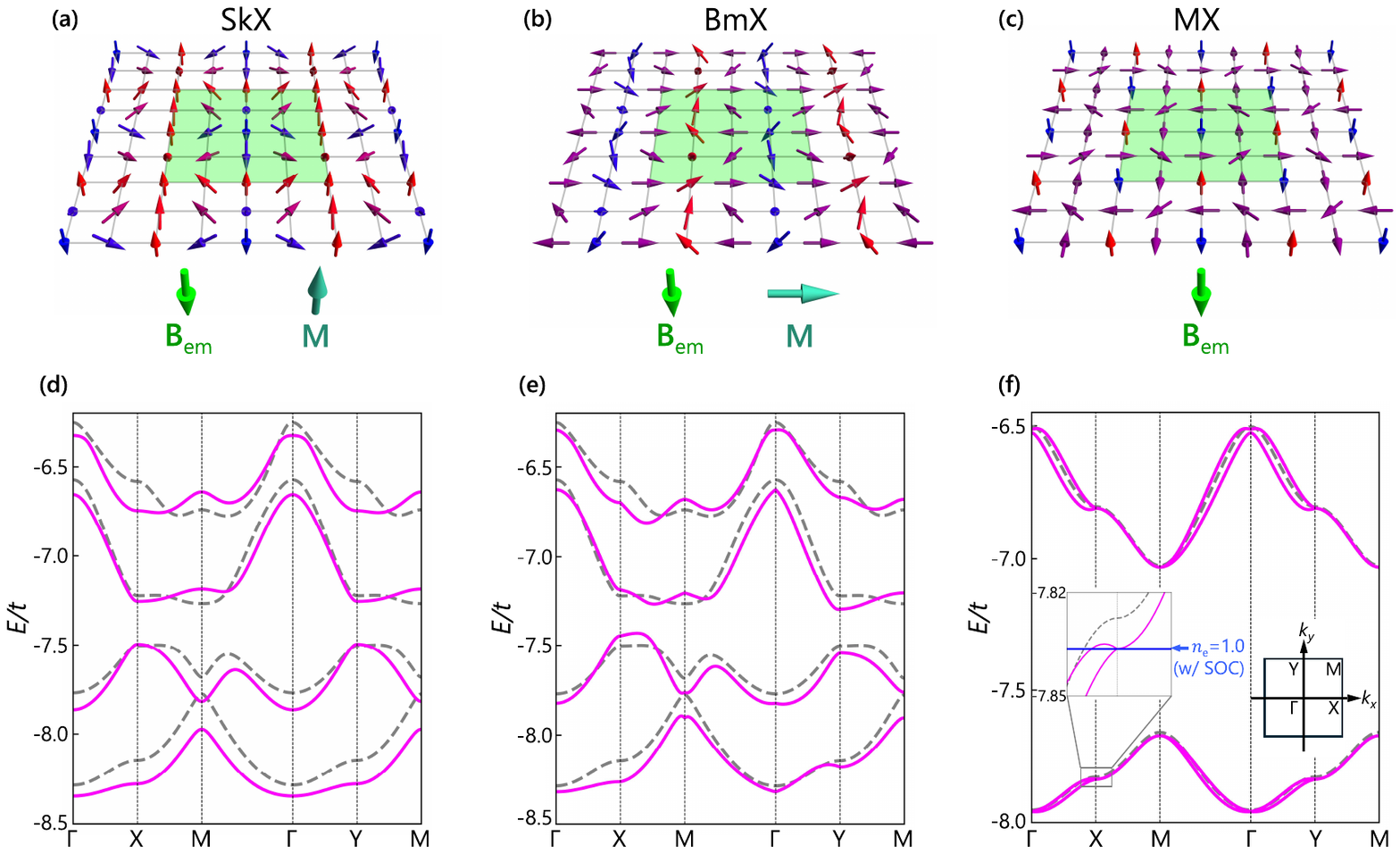}
    \caption{Spin configurations of three topological spin textures and corresponding electronic band structures. (a)--(c) Schematic illustrations of the spin configurations on a square lattice for (a) SkX ($\gamma=0$), (b) BmX ($\gamma=0$), and (c) MX ($\gamma=\pi/2$). The green-shaded squares represent the magnetic unit cell, which consists of $4 \times 4$ sites. The directions of the emergent magnetic field $\mathbf{B}_{\mathrm{em}}$ and the net magnetization $\mathbf{M}$ are also shown. (d)--(f) Electronic band structures on the high-symmetry lines (see the right inset) for (d) SkX, (e) BmX, and (f) MX. The model has 32 bands in total, and the four lowest-energy ones are shown. The gray dashed lines represent the results without SOC, while the magenta lines show the results with SOC, where the strength of SOC is set to $\lambda_{\mathrm{R}}=0.1t$. In (f), the left inset shows an enlarged view around the $\mathrm{X}$ point. The blue horizontal line indicates the Fermi energy at $n_\mathrm{e} = 1$ in the presence of SOC, which crosses the bands at the $\mathrm{X}$ point.}
    \label{fig:spin_texture_bands}
\end{figure*}

\subsection{Topological spin textures}
We consider three types of two-dimensional topological spin textures formed by the localized spins on the square lattice: SkX, BmX, and MX, as shown in Figs.~\hyperlink{fig:spin_texture_bands}{\ref{fig:spin_texture_bands}(a)}, \hyperlink{fig:spin_texture_bands}{\ref{fig:spin_texture_bands}(b)}, and \hyperlink{fig:spin_texture_bands}{\ref{fig:spin_texture_bands}(c)}, respectively. Each spin texture is constructed as a double-$Q$ spin structure, namely, a superposition of magnetic modulations with two orthogonal magnetic modulation vectors, $\mathbf{Q}_\mu=\frac{2\pi}{\lambda}\hat{\bm{\mu}} \ ({\mu}=x,y),$
where $\lambda$ represents the magnetic period and $\hat{\bm{\mu}}$ is the unit vector along the $\mu$ direction.

 The SkX can be regarded as a periodic arrangement of skyrmions. An individual skyrmion is characterized by three quantities: the vorticity $m$, distinguishing vortex from antivortex, the polarity $p$, indicating the relative orientation between the core and the outer spins, and the helicity $\gamma$, defined as the offset angle of the in-plane spin winding. In this paper, we focus on the SkX composed of skyrmions with $m=1$ and $p=1$. Its spin configuration is given by
\begin{equation}
 \mathbf{S}^{(\mathrm{SkX})}_i \propto\begin{pmatrix}
 \cos{\gamma}\sin{(\mathbf{Q}_{x}\cdot \mathbf{r}_i)}-\sin{\gamma}\sin{(\mathbf{Q}_{y}\cdot \mathbf{r}_i)} \\ \sin{\gamma}\sin{(\mathbf{Q}_{x}\cdot \mathbf{r}_i)}+\cos{\gamma}\sin{(\mathbf{Q}_{y}\cdot \mathbf{r}_i)} \\ M-\cos{(\mathbf{Q}_{x}\cdot \mathbf{r}_i)}-\cos{(\mathbf{Q}_{y}\cdot \mathbf{r}_i)}
\end{pmatrix},\label{eq:SkX_config}
\end{equation}
 where $M$ represents a uniform magnetization component that produces a net magnetization $\mathbf{M}$ in the out-of-plane direction. The noncoplanar spin structure of the SkX gives rise to an out-of-plane emergent magnetic field $\mathbf{B}_{\mathrm{em}}$ acting on itinerant electrons. In a two-dimensional spin texture, this field generally possesses only an out-of-plane component, which is given for each site $i$ by 
 \begin{align}
 B^z_{\mathrm{em}, i} = \frac{e}{2\hbar} \lbrace &\Omega(\mathbf{r}_i,\mathbf{r}_i+\mathbf{a}_1,\mathbf{r}_i+\mathbf{a}_2) \nonumber\\ &+\Omega(\mathbf{r}_i+\mathbf{a}_2,\mathbf{r}_i+\mathbf{a}_1,\mathbf{r}_i+\mathbf{a}_1+\mathbf{a}_2)\rbrace,
 \end{align}
 where $e$ denotes the elementary charge, $\hbar$ the reduced Planck constant, and $\mathbf{a}_1=(1,0,0)$ and $\mathbf{a}_2=(0,1,0)$ the primitive translation vectors of the square lattice, and $\Omega(\mathbf{r}_i,\mathbf{r}_j,\mathbf{r}_k)$ represents the solid angle subtended by three spins at sites $i,j,k$:
 \begin{align}
 \Omega(\mathbf{r}_i,&\mathbf{r}_j,\mathbf{r}_k) \nonumber \\
 &= 2\arctan\left[
 \frac{\mathbf{S}_i \cdot (\mathbf{S}_j \times \mathbf{S}_k)}{1 + \mathbf{S}_i\cdot\mathbf{S}_j + \mathbf{S}_j\cdot\mathbf{S}_k + \mathbf{S}_k\cdot\mathbf{S}_i}
 \right]
 \end{align}
 The spatial summation of this emergent magnetic field corresponds to the skyrmion number $N_{\mathrm{sk}}$, which characterizes the topology of this spin texture, as
 \begin{equation}
 N_{\mathrm{sk}}=\frac{\hbar}{2\pi e}\sum_{i\in \mathrm{mag. u. c.}}B^z_{\mathrm{em}, i} ,
 \end{equation}
where the summation is taken over the magnetic unit cell.
 In the present SkX in Eq.~\hyperlink{eq:SkX_config}{(\ref{eq:SkX_config})}, the skyrmion number is $N_{\mathrm{sk}}=-1$.
 
 The BmX is a periodic arrangement of bimerons. A bimeron is a bound pair of two merons \cite{Bm1,Bm2,Bm3,Bm4}. A meron is a spin texture characterized by an in-plane spin winding and an out-of-plane magnetization, carrying a skyrmion number $N_{\mathrm{sk}}=-1/2$ \cite{Meron1,Meron2,Meron3}. Similar to skyrmions, merons can be characterized by their vorticity, polarity, and helicity. A bimeron is formed by pairing two merons with opposite vorticity and polarity on a background of in-plane magnetization, resulting in a total skyrmion number $N_{\mathrm{sk}}=-1$. It can also be regarded as a skyrmion in which all spins are uniformly rotated by 90\textdegree\ so that the net magnetization lies in the plane \cite{Bm4}.  Accordingly, the BmX serves as the in-plane counterpart of the SkX. Figure \hyperlink{fig:spin_texture_bands}{\ref{fig:spin_texture_bands}(b)} represents the case with the net magnetization along the $x$ direction. The corresponding spin configuration is given by
\begin{equation}
     \mathbf{S}^{(\mathrm{BmX})}_i \propto\begin{pmatrix}
M-\cos{(\mathbf{Q}_{x}\cdot \mathbf{r}_i)}-\cos{(\mathbf{Q}_{y}\cdot \mathbf{r}_i)} \\ \sin{\gamma}\sin{(\mathbf{Q}_{x}\cdot \mathbf{r}_i)}+\cos{\gamma}\sin{(\mathbf{Q}_{y}\cdot \mathbf{r}_i)} \\ -\cos{\gamma}\sin{(\mathbf{Q}_{x}\cdot \mathbf{r}_i)}+\sin{\gamma}\sin{(\mathbf{Q}_{y}\cdot \mathbf{r}_i)} 
\end{pmatrix}.
\end{equation}
 Despite the different magnetization direction, the BmX shares the same skyrmion number $N_{\mathrm{sk}} = -1$ with the SkX, and therefore also exhibits a net out-of-plane emergent magnetic field.

 The MX is a periodic arrangement of merons \cite{Zhentao-MX,Hayami-MX}, whose spin configuration is given by 
 \begin{equation}
   \mathbf{S}^{(\mathrm{MX})}_i \propto\begin{pmatrix}
\cos{\gamma}\sin{(\mathbf{Q}_{x}\cdot \mathbf{r}_i)}-\sin{\gamma}\sin{(\mathbf{Q}_{y}\cdot \mathbf{r}_i)} \\
\sin{\gamma}\sin{(\mathbf{Q}_{x}\cdot \mathbf{r}_i)}+\cos{\gamma}\sin{(\mathbf{Q}_{y}\cdot \mathbf{r}_i)}\\
-\cos{(\mathbf{Q}_{x}\cdot \mathbf{r}_i)}\cos{(\mathbf{Q}_{y}\cdot \mathbf{r}_i)}
\end{pmatrix}.
 \end{equation}
In contrast to SkX and BmX, the MX exhibits no net magnetization. Within the magnetic unit cell, four merons with different vorticity and polarity are arranged, each carrying the skyrmion number of $N_{\mathrm{sk}}=-1/2$.
Consequently, the total skyrmion number per magnetic unit cell becomes $N_{\mathrm{sk}}=-2$, and thus, this case also accompanies a net emergent magnetic field along the out-of-plane direction. 

  In the following calculations, we take the magnetic period of each spin texture as $\lambda=4$, so that the magnetic unit cell forms a $4\times4$ square, as indicated by the green-shaded squares in Figs.~\hyperlink{fig:spin_texture_bands}{\ref{fig:spin_texture_bands}(a)}, \hyperlink{fig:spin_texture_bands}{\ref{fig:spin_texture_bands}(b)}, and \hyperlink{fig:spin_texture_bands}{\ref{fig:spin_texture_bands}(c)}. The magnetization parameter is set to $M=0.3$ for both SkX and BmX. In the main text, we focus on the case $\gamma = 0$ for SkX and BmX, called N\'{e}el-type, and the case $\gamma = \pi/2$ for MX, called Bloch-type. Results for other helicities are provided in  Appendix \ref{apx:Results for other helicities}.
  
\section{Method}
\label{sec:Method}
\subsection{Electronic property}
 We construct the Bloch Hamiltonian in the magnetic Brillouin zone for the model in Eq.~\hyperlink{eq:Hamiltonian}{(\ref{eq:Hamiltonian})} and obtain the electronic states by exact diagonalization. The resulting eigenvalues and eigenvectors serve as the basis for calculating band structures and transport properties under each topological spin texture. These procedures are performed via the Fourier transformation $\hat{c}_{i\alpha}=\frac{1}{\sqrt{N}}\sum_{\mathbf{k}}\hat{c}_{\mathbf{k}l\alpha}\exp(\mathrm{i}\mathbf{k}\cdot\mathbf{r}_i)$, where $N$ denotes the total number of magnetic unit cells in the system, $\mathbf{k}$ is the wave number in the magnetic Brillouin zone, and $l$ is the sublattice index of site $i$. The Hamiltonian is thereby rewritten in momentum space as $\hat{\mathcal{H}}=\sum_{\mathbf{k}}\vec{c}^{\dag}_{\mathbf{k}}H_{\mathbf{k}}\vec{c}_{\mathbf{k}}$, where $\vec{c}_{\mathbf{k}}$ denotes a spinor that contains all sublattice and spin degrees of freedom. From the energy eigenstates of the Bloch Hamiltonian $\vec{c}^{\dag}_{\mathbf{k}}H_{\mathbf{k}}\vec{c}_{\mathbf{k}}$, we calculate the spin expectation value $s_{a}(\mathbf{k};n)$ for $n$th band at momentum $\mathbf{k}$, with the index $a=x,y,z$ denoting the spin component, as
 \begin{equation}
     s_{a}(\mathbf{k};n)=\mel{n\mathbf{k}}{\hat{s}^{a}_{\mathbf{k}}}{n\mathbf{k}},
 \end{equation}
 where $\ket{n\mathbf{k}}$ is the Bloch eigenstate and $\hat{s}^{a}_{\mathbf{k}}=\frac{\hbar}{2}\sum_{l,\alpha,\beta}\hat{c}^{\dag}_{\mathbf{k}l\alpha}\sigma^a_{\alpha\beta}\hat{c}_{\mathbf{k}l\beta}$ is the spin operator acting on the states with momentum $\mathbf{k}$. These calculations are performed by varying the electron filling $n_{\mathrm{e}}$, which represents the average number of itinerant electrons per magnetic unit cell:
 
 \begin{equation}
 n_{\mathrm{e}}=\frac{1}{N}\sum_{n,\mathbf{k}}f(\varepsilon_{n\mathbf{k}})=\frac{1}{N}\sum_{n,\mathbf{k}}\frac{1}{e^{(\varepsilon_{n\mathbf{k}}-\varepsilon_{\mathrm{F}})/k_{\mathrm{B}}T}+1},
 \end{equation}
 where $f(\varepsilon_{n\mathbf{k}})$ denotes the Fermi--Dirac distribution function, $\varepsilon_{n\mathbf{k}}$ the energy of the Bloch eigenstate $\ket{{n\mathbf{k}}}$, $k_{\mathrm{B}}$ the Boltzmann constant, $T$ the temperature, and $\varepsilon_{\mathrm{F}}$ the Fermi energy.
 
 In the following calculations, we perform the exact diagonalization of the Bloch Hamiltonian on a $300^2$ mesh in the magnetic Brillouin zone. Although the electron filling $n_{\mathrm{e}}$ can range up to $32$ in the present model for $\lambda=4$, we restrict our analysis to the low-filling region $0\leq n_{\mathrm{e}}\leq 4$.

\subsection{Transport property}
 We address both charge and spin transport by evaluating the conductivity tensor in each channel within the linear response. The linear charge conductivity tensor describes an electric current $J_{\mu}$ in the $\mu$ direction $(\mu=x,y)$ induced by an applied electric field $E_{\nu}$ in the $\nu$ direction $(\nu=x,y)$ as 
 \begin{equation}
     J_{\mu}=(\sigma_{\mu\nu}+\sigma^{(\tau)}_{\mu\nu})E_\nu,
 \end{equation}
  where $\sigma_{\mu\nu}$ describes the intrinsic contribution independent of scattering of electrons, while $\sigma^{(\tau)}_{\mu\nu}$ represents the dissipative contribution from the Fermi surface, proportional to the transport relaxation time $\tau$. The electric current operator $\hat{J}_{\mu}$ is defined as 
  \begin{equation}
      \hat{J}_{\mu}=-e\hat{v}_\mu,
  \end{equation}
  where the velocity operator in the $\mu$ direction is given by 
  \begin{equation}
  \hat{v}_\mu=\frac{1}{\hbar}\sum_{\mathbf{k}}\vec{c}^{\dag}_{\mathbf{k}}\frac{\partial H_{\mathbf{k}}}{\partial k_\mu}\vec{c}_{\mathbf{k}}.
  \end{equation}
  By using the Kubo formula, the intrinsic term is calculated via current-current correlations as
 \begin{equation}
     \sigma_{\mu\nu}=\dfrac{e^2}{2\pi h}\int d\mathbf{k} \sum_{n} f(\varepsilon_{n\mathbf{k}})B_{\mu\nu}(\mathbf{k};n),
 \end{equation}
 where $h$ denotes the Planck constant and $B_{\mu\nu}(\mathbf{k};n)$ is the Berry curvature of $n$th band, while the integration is performed over the entire magnetic Brillouin zone. The Berry curvature is evaluated as
 \begin{equation}
 \hypertarget{eq:Berry}{}
     B_{\mu\nu}(\mathbf{k};n)=2\hbar^2\sum_{m \neq n} \mathrm{Im}\left[\dfrac{\mel**{n\mathbf{k}}{\hat{v}_{\mu}}{m\mathbf{k}}\mel**{m\mathbf{k}}{\hat{v}_{\nu}}{n\mathbf{k}}}{(\varepsilon_{n\mathbf{k}}-\varepsilon_{m\mathbf{k}})^2}\right].
 \label{eq:Berry}
 \end{equation}
 Meanwhile, the dissipative term is computed by adopting the Boltzmann transport theory, without assuming any specific microscopic scattering mechanism. The expression is given by
 \begin{equation}
 \sigma^{(\tau)}_{\mu\nu}=\dfrac{e\tau}{h}\dfrac{1}{2\pi}\int d\mathbf{k} \sum_{n}\mel**{n\mathbf{k}}{\hat{J}_{\mu}}{n\mathbf{k}}\dfrac{\partial f(\varepsilon_{n\mathbf{k}})}{\partial k_{\nu}}.    
\end{equation}
 Here, the transport relaxation time $\tau$ is assumed to be independent of band index $n$ and wave number $\mathbf{k}$.

 For the spin transport properties, we introduce the spin current operator \cite{Sinova_Conv},
 \begin{equation}
     \hat{J}^{s_a}_{\mu}=\dfrac{1}{2} \lbrace \hat{v}_\mu, \hat{s}_a \rbrace,
 \end{equation}
where $\hat{s}_a=\sum_{\mathbf{k}}\hat{s}^{a}_{\mathbf{k}}$ is the $a$ component of the spin operator, and $\lbrace \ldots\rbrace$ denotes the anticommutator. This operator represents the flow of the spin angular momentum along the $\mu$ axis with polarization along the $a$ axis. Note that the definition of spin current in spin-nonconserving systems has been discussed extensively \cite{Rashba_SC,Proper_SC,Jin_2006,An2012,tamaya2024}, but a well-established formulation remains elusive. Since our model does not conserve each spin component of itinerant electrons, we adopt the above operator as an ad hoc ansatz to approximate spin transport. Using this definition, we evaluate the linear spin conductivity tensor, which describes a spin current $J^{s_a}_\mu$ induced by an applied electric field $E_\nu$, as
\begin{equation}
     J_{\mu}^{s_a}=(\sigma^{s_a}_{\mu\nu}+\sigma^{s_a(\tau)}_{\mu\nu})E_\nu,
 \end{equation}
 where $\sigma^{s_a}_{\mu\nu}$ and $\sigma^{s_a(\tau)}_{\mu\nu}$ represent intrinsic and dissipative contributions,  respectively, as in the charge transport.
The intrinsic term is calculated using the Kubo formula as
 \begin{equation}
     \sigma^{s_a}_{\mu\nu}=-\dfrac{e}{4\pi^2}\int d\mathbf{k} \sum_{n} f(\varepsilon_{n\mathbf{k}})B^{s_a}_{\mu\nu}(\mathbf{k};n),
 \end{equation}
 where the spin Berry curvature $B^{s_a}_{\mu\nu}$ is defined as
 \begin{equation}
 \hypertarget{eq:SpinBerry}{}
     B^{s_a}_{\mu\nu}(\mathbf{k};n)=2\hbar\sum_{m \neq n} \mathrm{Im}\left[\dfrac{\mel**{n\mathbf{k}}{\hat{J}^{s_a}_{\mu}}{m\mathbf{k}}\mel**{m\mathbf{k}}{\hat{v}_{\nu}}{n\mathbf{k}}}{(\varepsilon_{n\mathbf{k}}-\varepsilon_{m\mathbf{k}})^2}\right].
 \label{eq:SpinBerry}
 \end{equation}
 Meanwhile, the dissipative term is obtained from the Boltzmann transport equation as
 \begin{equation}
 \sigma^{s_a(\tau)}_{\mu\nu}=\dfrac{e\tau}{h}\dfrac{1}{2\pi}\int d\mathbf{k} \sum_{n}\mel**{n\mathbf{k}}{\hat{J}^{s_a}_{\mu}}{n\mathbf{k}}\dfrac{\partial f(\varepsilon_{n\mathbf{k}})}{\partial k_{\nu}}.    
 \end{equation}

\section{Results}

\begin{figure*}[htbp]
    \centering
    \hypertarget{fig:3Dbands}{}
    \includegraphics[width=\textwidth]{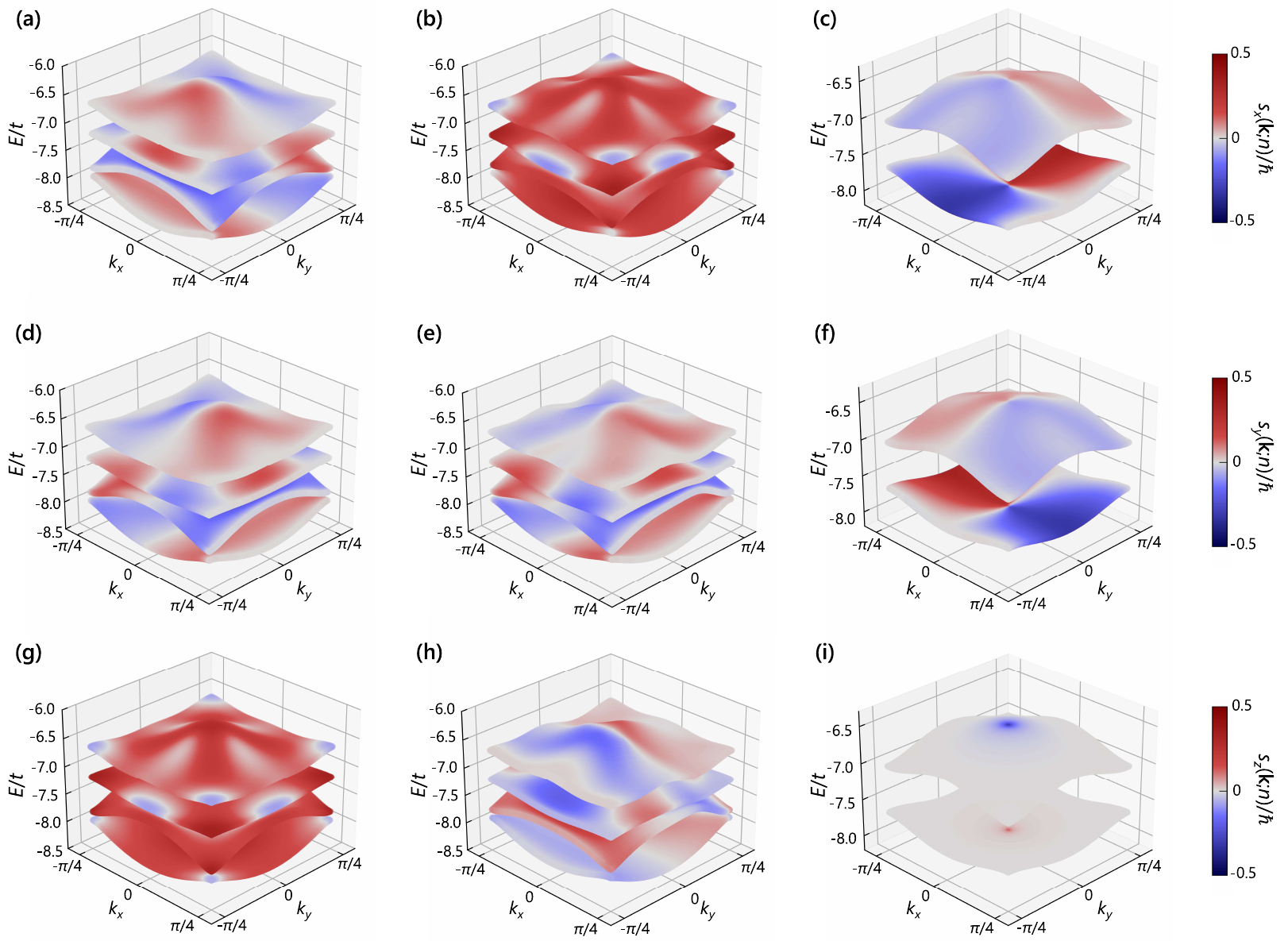}
    \caption{Spin-projected electronic band structures in the presence of SOC. The left, center, and right columns correspond to SkX, BmX, and MX, respectively. In each panel, the surface color indicates the spin expectation value $s_a(\mathbf{k};n)$, where $a=x,y, \mathrm{and}\ z$ for panels (a)--(c), (d)--(f), and (g)--(i), respectively.}
    \label{fig:3Dbands}
\end{figure*}

\subsection{Band structure and spin splitting}
\label{subsec: Bands}
 Before discussing the spin transport properties, here we investigate the electronic band structure and spin splittings under each spin texture. These features play a crucial role in shaping spin--momentum locking at the Fermi surfaces and the spin Berry curvature, which directly govern the spin current generation. In particular, the presence or absence of spin splittings and the symmetry of the band structures impose constraints on the form of the spin conductivity tensor.

 We first focus on the SkX and BmX cases.
 Figures~\hyperlink{fig:spin_texture_bands}{\ref{fig:spin_texture_bands}(d)} and \hyperlink{fig:spin_texture_bands}{\ref{fig:spin_texture_bands}(e)} show the electronic energy dispersions for SkX and BmX, respectively, along the high-symmetry lines of the magnetic Brillouin zone [see the right inset of Fig.~\hyperlink{fig:spin_texture_bands}{\ref{fig:spin_texture_bands}(f)}]. The gray dashed and magenta lines represent the results without and with SOC, respectively. Without SOC, the band structures for SkX and BmX are identical, whereas introducing SOC leads to clear differences between them. This indicates that the SOC accentuates the distinctions between SkX and BmX in the electronic structures. Note that in both cases, band gaps open between the lowest and second-lowest energy bands, as well as between the second and third-lowest energy bands, indicating that the system becomes insulating at electron fillings $n_{\mathrm{e}}=1$ and $2$.
 
 The differences between SkX and BmX are also highlighted in the momentum-space distribution of the spin expectation value of each band. Figure~\ref{fig:3Dbands} shows the results with SOC: the panels \hyperlink{fig:3Dbands}{(a)}, \hyperlink{fig:3Dbands}{(d)}, and \hyperlink{fig:3Dbands}{(g)} correspond to the $x$, $y$, and $z$ spin components for the SkX, respectively, while \hyperlink{fig:3Dbands}{(b)}, \hyperlink{fig:3Dbands}{(e)}, and \hyperlink{fig:3Dbands}{(h)} correspond to those for the BmX. For the SkX, the energy dispersions retain fourfold rotational symmetry, whereas the spin polarization of each band exhibits characteristic symmetry depending on the spin component. Specifically, $s_x(\mathbf{k};n)$ is symmetric with respect to $k_x=0$ and antisymmetric with respect to $k_y=0$ as shown in Fig.~\hyperlink{fig:3Dbands}{\ref{fig:3Dbands}(a)}, while $s_y(\mathbf{k};n)$ exhibits the opposite behavior as shown in Fig.~\hyperlink{fig:3Dbands}{\ref{fig:3Dbands}(d)}. Due to these symmetry characteristics, both $s_x$ and $s_y$ components cancel out upon momentum integration, resulting in no net in-plane polarization. By contrast, $s_z({\bf k};n)$ exhibits fourfold symmetry and gives a nonzero net polarization. Consequently, the spins of itinerant electrons are overall polarized along the net magnetization of the SkX. Similarly, the BmX generates overall spin polarization along its magnetization direction ($x$ direction), as indicated by the strong intensity of $s_x({\bf k};n)$ in Fig.~\hyperlink{fig:3Dbands}{\ref{fig:3Dbands}(b)}. However, unlike the SkX case, the fourfold rotational symmetry of the energy dispersions is broken, while they retain symmetry with respect to $k_x=0$. The spin expectation values also exhibit reduced symmetry: $s_y(\mathbf{k};n)$ is antisymmetric and $s_z(\mathbf{k};n)$ is symmetric with respect to $k_x=0$, yet neither shows a clear symmetry with respect to $k_y=0$, as shown in Figs.~\hyperlink{fig:3Dbands}{\ref{fig:3Dbands}(e)} and \hyperlink{fig:3Dbands}{\ref{fig:3Dbands}(h)}. Thus, the BmX exhibits a lowering of symmetry compared with the SkX under SOC, which leads to different spin transport properties, as discussed in Sec.~\ref {subsec: transport}. A detailed symmetry analysis will be given in Sec.~\ref{subsec: symmetry}.
 
 Now we turn to the MX. Figure~\hyperlink{fig:spin_texture_bands}{\ref{fig:spin_texture_bands}(f)} shows the electronic band structure for the MX without and with SOC. In this case, band gaps open between the second and third, and between the fourth and fifth-lowest energy bands, leading to insulating states at $n_{\mathrm{e}}=2$ and $4$. Without SOC, unlike the SkX and BmX cases, there is no spin splitting in the band structure, reflecting the absence of the net magnetization in the MX; each band is doubly degenerate. By introducing SOC, however, spin splitting occurs, while the degeneracy remains along the magnetic Brillouin zone boundary, namely the $\mathrm{XM}$ and $\mathrm{YM}$ lines.
 The spin-projected bands with SOC are shown in Figs.~\hyperlink{fig:3Dbands}{\ref{fig:3Dbands}(c)}, \hyperlink{fig:3Dbands}{\ref{fig:3Dbands}(f)}, and \hyperlink{fig:3Dbands}{\ref{fig:3Dbands}(i)}. They exhibit the same symmetry as for the SkX case. As shown in Figs.~\hyperlink{fig:3Dbands}{\ref{fig:3Dbands}(c)} and \hyperlink{fig:3Dbands}{\ref{fig:3Dbands}(f)}, the $s_x$ and $s_y$ components are antisymmetric with respect to $k_y=0$ and $k_x=0$, respectively, therefore the net in-plane spin polarization vanishes. In contrast, despite the absence of the net magnetization in the MX, the net out-of-plane spin polarization does not cancel. Notably, $s_z(\mathbf{k};n)$ exhibits fourfold rotational symmetry around the $z$ direction, which becomes pronounced at the $\mathrm{\Gamma}$ point, as shown in Fig.~\hyperlink{fig:3Dbands}{\ref{fig:3Dbands}(i)}. The emergence of momentum-dependent spin splitting due to SOC implies that the SOC plays a crucial role in spin transport properties for the MX, as discussed in the following sections.
 
\begin{figure*}[htbp]
    \centering
    \hypertarget{fig:Spincond}{}
    \includegraphics[width=\textwidth]{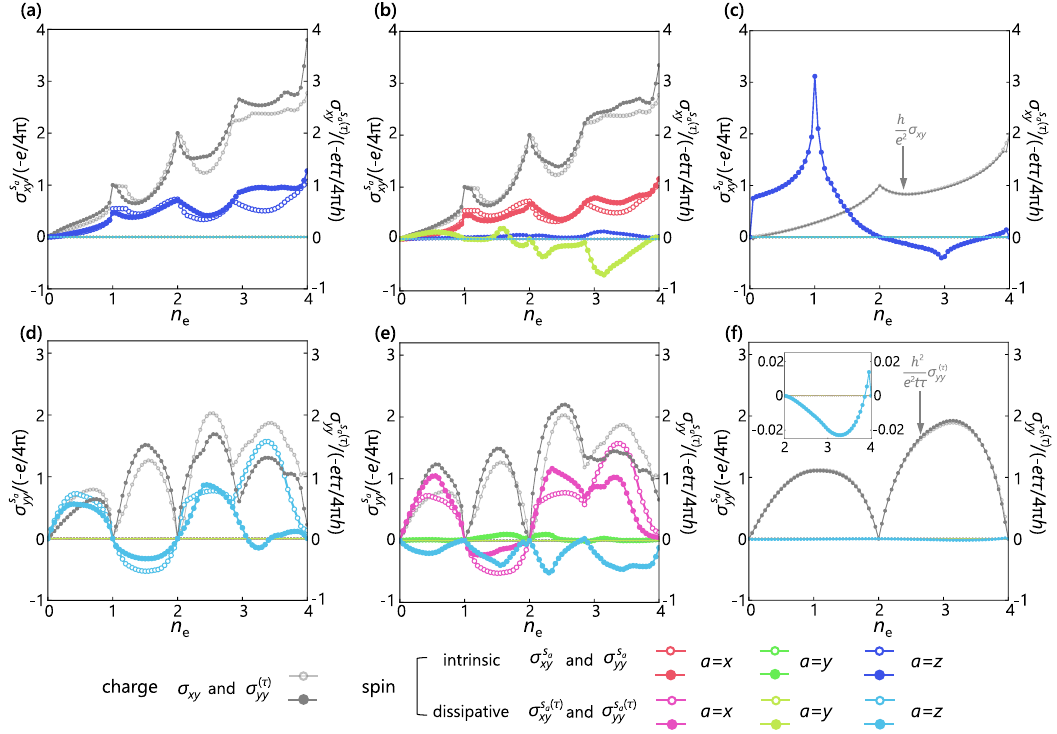}
    \caption{Electron filling dependencies of the charge and spin conductivities for three spin textures. The left, center, and right columns correspond to SkX, BmX, and MX, respectively: (a)--(c) charge Hall conductivity $\sigma_{xy}$ (intrinsic, plotted as the dimensionless quantity $h\sigma_{xy}/e^2$), and the spin Hall conductivities $\sigma^{s_a}_{xy}$ (intrinsic, left axes) and $\sigma^{s_a(\tau)}_{xy}$ (dissipative, right axes), and (d)--(f) longitudinal charge conductivity $\sigma^{(\tau)}_{yy}$ (dissipative, plotted as the dimensionless quantity $h^2\sigma^{(\tau)}_{yy}/e^2t\tau$), and spin conductivities $\sigma^{s_a}_{yy}$ (intrinsic, left axes) and $\sigma^{s_a(\tau)}_{yy}$ (dissipative, right axes). In each panel, the charge conductivity is plotted in gray, while the spin conductivities are plotted in different colors according to the spin polarization direction $a=x,y,z$ and the type of contribution (intrinsic or dissipative), as shown in the legend below the panels. For both the charge and spin conductivities, the open circles indicate the results without SOC, and the filled circles represent the results with SOC. The inset in (f) shows an enlarged view of $\sigma^{s_z(\tau)}_{yy}$ for $2\leq n_{\mathrm{e}}\leq4$.}
    \label{fig:Spincond}
\end{figure*}

\begin{figure*}[htbp]
    \centering
    \hypertarget{fig:Berry}{}
    \includegraphics[width=1\textwidth]{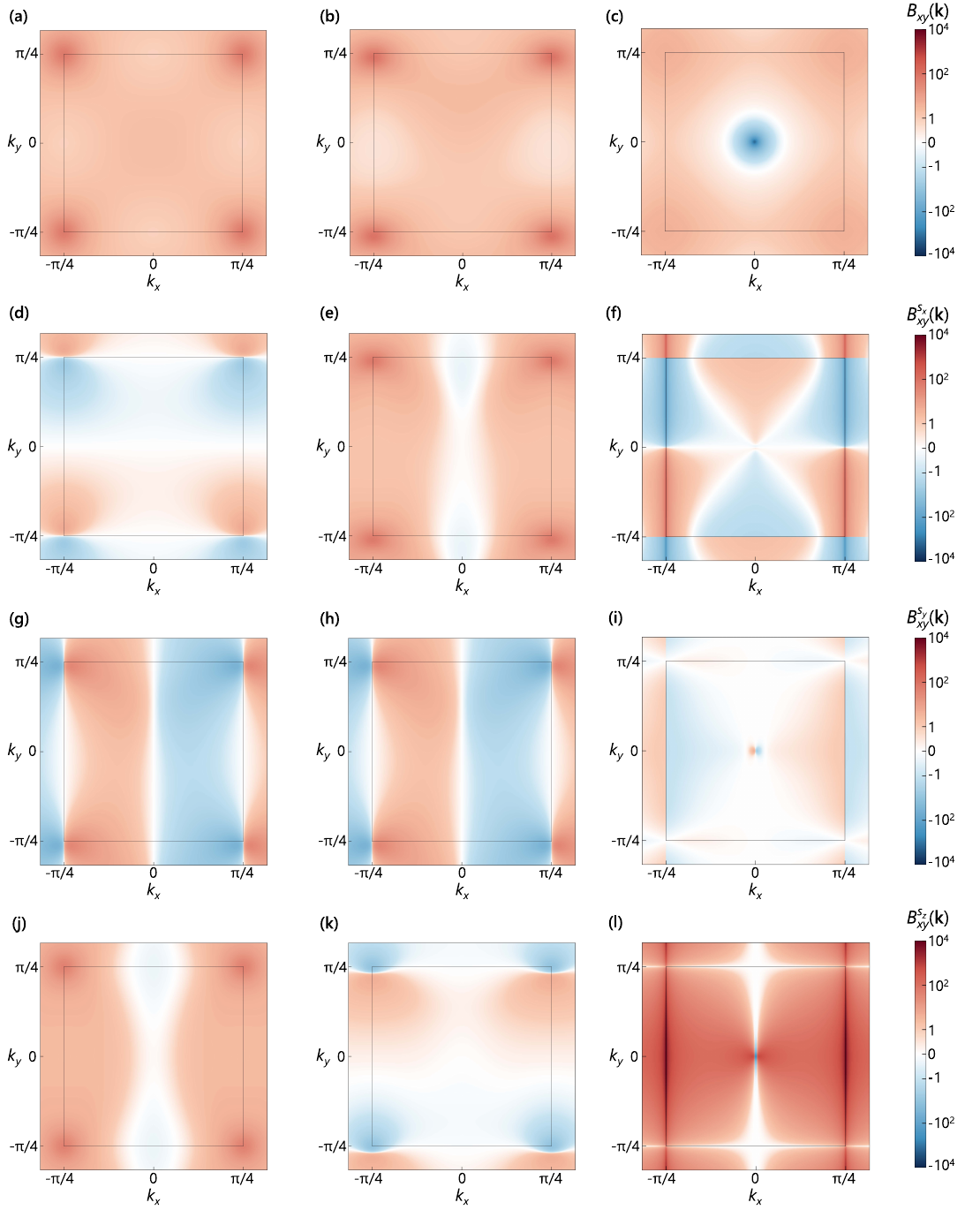}
    \caption{Berry curvature and spin Berry curvature for each spin texture in the presence of SOC. The left, center, and right columns correspond to SkX, BmX, and MX, respectively. (a)--(c) Momentum-space distributions of the Berry curvature $B_{xy}$ [Eq.~\hyperref[eq:Berry]{(\ref*{eq:Berry})}] for the lowest-energy band. (d)--(l) Spin Berry curvature $B^{s_a}_{xy}$ [Eq.~\hyperref[eq:SpinBerry]{(\ref*{eq:SpinBerry})}] for the lowest-energy band for the spin components: (d)--(f) $a=x$, (g)--(i) $a=y$, and (j)--(l) $a=z$. The black squares indicate the magnetic Brillouin zone boundary.}
    \label{fig:Berry}
\end{figure*}
\subsection{Charge and spin transport}
\label{subsec: transport}
\subsubsection{SkX and BmX}
 In this section, we investigate the charge and spin conductivities for the SkX and the BmX at zero temperature ($T=0$), both with and without SOC. 
 While the previous study mainly addressed the intrinsic out-of-plane spin Hall conductivity $\sigma^{s_z}_{xy}$ in noncoplanar spin textures, including SkX and BmX in the absence of SOC \cite{TOHE}, we here provide a comprehensive analysis including the in-plane spin polarization components, longitudinal transport, and dissipative contributions from Fermi surfaces, with particular attention to the effects of SOC.
 
 We begin with the charge conductivity.
 The electrical Hall conductivities for SkX and BmX are shown by the gray symbols in Figs.~\hyperlink{fig:Spincond}{\ref{fig:Spincond}(a)} and \hyperlink{fig:Spincond}{\ref{fig:Spincond}(b)}, respectively. These transverse responses originate from the topological Hall effect induced by the emergent magnetic field of each spin texture, with an additional contribution from the anomalous Hall effect when the SOC is present. In the absence of SOC, the Hall conductivities for SkX and BmX are identical. The Berry curvature distributions in momentum space in the presence of SOC are shown in Figs.~\hyperlink{fig:Berry}{\ref{fig:Berry}(a)} and \hyperlink{fig:Berry}{\ref{fig:Berry}(b)}, indicating that it is strongly concentrated around the $\mathrm{M}$ points. In the insulating states at $n_{\mathrm{e}}=1$ and $2$, the Hall conductivity is quantized to integer multiples of $e^2/h$, known as the quantized topological Hall effect \cite{QTHE}, reflecting the nontrivial topology of the spin texture. The longitudinal charge conductivities are shown by the gray lines in Figs.~\hyperlink{fig:Spincond}{\ref{fig:Spincond}(d)} and \hyperlink{fig:Spincond}{\ref{fig:Spincond}(e)}. In the absence of SOC, they are identical between SkX and BmX, as in the case of the Hall conductivity.

 We next examine the spin conductivities for SkX and BmX. The results without SOC are shown by open colored circles in Figs.~\hyperlink{fig:Spincond}{\ref{fig:Spincond}(a)}, \hyperlink{fig:Spincond}{\ref{fig:Spincond}(b)}, \hyperlink{fig:Spincond}{\ref{fig:Spincond}(d)}, and \hyperlink{fig:Spincond}{\ref{fig:Spincond}(e)}. For each texture, only two components become nonzero: the intrinsic transverse and dissipative longitudinal components with spin polarization along the magnetization direction, namely $\sigma^{s_z}_{xy}$ and $\sigma^{s_z(\tau)}_{yy}$ for the SkX, and $\sigma^{s_x}_{xy}$ and $\sigma^{s_x(\tau)}_{yy}$ for the BmX. This result can be understood by noting that the itinerant electrons are overall polarized along the magnetization direction, and these spin-polarized carriers contribute to the topological spin Hall response through the emergent magnetic field in the transverse component and to the Drude response in the longitudinal component. In other words, the transverse response corresponds to the conventional topological spin Hall effect widely studied in skyrmions \cite{TSHE_sk1,TSHE_sk2,TOHE}, whereas the longitudinal response simply reflects the Drude conduction by spin-polarized electrons in a ferromagnet \cite{Mott1,Mott2,TwoCurrmodel1,TwoCurrmodel2,SpolC1,SpolC2,STT3,STT4,STT5}. Although the polarization direction for the nonzero spin current differs between SkX and BmX, the magnitudes of the corresponding spin components remain identical. Hence, the transport properties for the SkX and BmX in the absence of SOC have one-to-one correspondence via the rotation of the spin configuration.

 We now turn to the spin conductivities in the presence of SOC, shown by the filled colored circles in Figs.~\hyperlink{fig:Spincond}{\ref{fig:Spincond}(a)}, \hyperlink{fig:Spincond}{\ref{fig:Spincond}(b)}, \hyperlink{fig:Spincond}{\ref{fig:Spincond}(d)}, and \hyperlink{fig:Spincond}{\ref{fig:Spincond}(e)}.
 The corresponding distributions of the spin Berry curvature, which contribute to the intrinsic spin Hall conductivities, are shown in the left and center columns in Fig.~\hyperlink{fig:Berry}{\ref{fig:Berry}}. For the SkX, the overall behavior remains qualitatively the same as the case without SOC. Only the components polarized along the magnetization direction, namely $\sigma^{s_z}_{xy}$ and $\sigma^{s_z(\tau)}_{yy}$, become nonzero, while the other components remain zero. In contrast, the spin conductivities for the BmX exhibit qualitatively different behavior compared to the case without SOC. In addition to the components with polarization along the magnetization direction, other components, intrinsic $\sigma^{s_z}_{xy}$ and $\sigma^{s_y}_{yy}$ and dissipative $\sigma^{s_y(\tau)}_{xy}$ and $\sigma^{s_z(\tau)}_{yy}$, become nonzero. The additional $\sigma^{s_z}_{xy}$ can be understood from the spin Berry curvature distribution. As shown in Figs.~\hyperlink{fig:Berry}{\ref{fig:Berry}(h)} and \hyperlink{fig:Berry}{\ref{fig:Berry}(k)}, $B^{s_z}_{xy}$ lacks antisymmetry and thus does not cancel upon momentum integration, giving rise to nonzero $\sigma^{s_z}_{xy}$, while $B^{s_y}_{xy}$ remains antisymmetric with respect to $k_x = 0$ and cancels out, yielding $\sigma^{s_y}_{xy}=0$. Meanwhile, $\sigma_{yy}^{s_y}$ stems from the nonzero values of $B_{yy}^{s_y}$ (not shown), whose physical interpretation remains elusive; it may originate from the ambiguity in the definition of the spin current operator in spin-nonconserving systems. The additional nonzero dissipative terms, $\sigma^{s_y(\tau)}_{xy}$ and $\sigma^{s_z(\tau)}_{yy}$, can be understood as a consequence of the reduced symmetry of the Fermi surfaces associated with the modulated band structures in  Fig.~\hyperlink{fig:3Dbands}{\ref{fig:3Dbands}}.

 Thus, the introduction of SOC makes the spin transport properties for the BmX distinct from those for the SkX. The reduced symmetry of the electronic structure for the BmX under SOC, discussed in Sec.~\ref{subsec: Bands}, leads to the lowering of symmetry in both the band structure and the spin Berry curvature. This demonstrates that the intrinsically lower magnetic symmetry of the BmX compared with the SkX becomes evident in spin transport properties via SOC. The corresponding group theoretical analysis will be given in Sec.~\ref{subsec: symmetry}.

\subsubsection{MX}
 We turn to the results for the MX. Similarly to the SkX and BmX, we observe a nonzero electrical Hall conductivity, as shown in Fig.~\hyperlink{fig:Spincond}{\ref{fig:Spincond}(c)}. The quantized topological Hall effect also occurs at the insulating fillings $n_{\mathrm{e}}=2 \mathrm{\ and\ } 4$ \cite{Zhentao-MX}. In addition, we observe a nonzero longitudinal charge conductivity for the other metallic fillings, as shown in Fig.~\hyperlink{fig:Spincond}{\ref{fig:Spincond}(f)}.

 In the absence of SOC, all components of the spin conductivity vanish, as shown in Figs.~\hyperlink{fig:Spincond}{\ref{fig:Spincond}(c)} and \hyperlink{fig:Spincond}{\ref{fig:Spincond}(f)}. This result can be attributed to the lack of spin splitting in the band structure discussed in Sec.~\ref{subsec: Bands}. In contrast, when the SOC is switched on, nonzero components of the spin conductivity appear in the same manner as those for the SkX, namely the intrinsic $\sigma^{s_z}_{xy}$ and the dissipative $\sigma^{s_z(\tau)}_{yy}$. This similarity stems from the fact that the symmetry of the electronic structure for the MX resembles that for the SkX, as indicated in Fig.~\hyperlink{fig:3Dbands}{\ref{fig:3Dbands}}. Indeed, as shown in Figs.~\hyperlink{fig:Berry}{\ref{fig:Berry}(f)}, \hyperlink{fig:Berry}{\ref{fig:Berry}(i)}, and \hyperlink{fig:Berry}{\ref{fig:Berry}(l)}, the spin Berry curvature for the MX shares the same symmetry characteristics as that for the SkX, such as the symmetries with respect to $k_x= 0$ and $k_y=0$. The dissipative $\sigma^{s_z(\tau)}_{yy}$ is small but nonzero, as shown in the inset of Fig.~\hyperlink{fig:Spincond}{\ref{fig:Spincond}(f)}, which is consistent with the weak spin polarization along the $z$ direction shown in Fig.~\hyperlink{fig:3Dbands}{\ref{fig:3Dbands}{(i)}}. 
 
 A crucial difference from the SkX is that the intrinsic $\sigma^{s_z}_{xy}$ exhibits pronounced peaks at $n_{\mathrm{e}}=1$ and $3$, as shown in Fig.~\hyperlink{fig:Spincond}{\ref{fig:Spincond}(c)}. These peaks arise from enhanced spin Berry curvature at these commensurate fillings. We explain this mechanism, focusing on the peak at $n_{\mathrm{e}}=1$. As discussed in Sec.~\ref{subsec: Bands}, the band structure for the MX with SOC has spin degeneracy along the $\mathrm{XM}$ and $\mathrm{YM}$ lines [Fig.~\hyperlink{fig:spin_texture_bands}{\ref{fig:spin_texture_bands}(f)}]. Since the energy difference between different bands appears in the denominator of the expression for the spin Berry curvature [Eq.~\hyperlink{eq:SpinBerry}{(\ref{eq:SpinBerry})}], it is likely to have large contributions near such band degeneracy. In fact, the momentum-space distribution of the spin Berry curvature $B^{s_z}_{xy}$ of the lowest-energy band exhibits divergent behavior near the $\mathrm{XM}$ lines, as shown in Fig.~\hyperlink{fig:Berry}{\ref{fig:Berry}(i)}. At $n_{\mathrm{e}}=1$, the Fermi energy crosses the bands near the X point [Fig.~\hyperlink{fig:spin_texture_bands}{\ref{fig:spin_texture_bands}(f)}], where contributions from the enhanced spin Berry curvature dominate, leading to the pronounced peak of the spin Hall conductivity.
 
 Thus, even though the MX has no net magnetization, a sizable spin Hall conductivity is realized in the presence of SOC. The magnitude of $\sigma^{s_z}_{xy}$ can exceed the value for a nonmagnetic 2D electron system with Rashba SOC, $e/8\pi$ \cite{Sinova_Conv}, by more than a factor of six. Since the MX does not generate spin current in the absence of SOC, this pronounced spin Hall response arises from the cooperative effect between the SOC and the topological spin texture. In particular, the distinctive symmetry of MX, which preserves spin degeneracy at the Brillouin zone boundary, plays a crucial role in enhancing the response.

\begin{figure}[htbp]
    \centering
    \includegraphics[width=\linewidth]{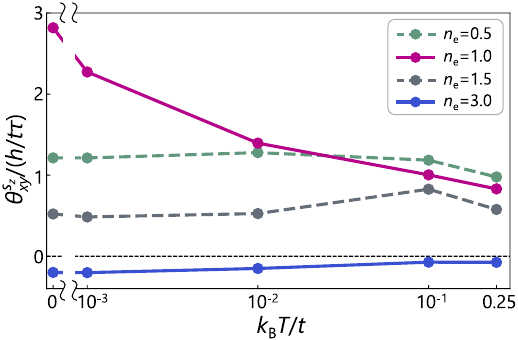}
    \caption{Temperature dependence of the spin Hall angle $\theta_{xy}^{s_z}$ for the MX at several electron fillings $n_{\mathrm{e}}$ in the presence of SOC. }
    \label{fig:MX_SHA}
\end{figure}

 Given these enhancements, we evaluate the efficiency of spin current generation. For this purpose, we calculate the spin Hall angle, which represents a ratio of charge-to-spin conversion, defined by
 \begin{equation}
     \theta^{s_z}_{xy}=-\frac{2e}{\hbar}\frac{\sigma^{s_z}_{xy}}{\sigma^{(\tau)}_{yy}}.
 \end{equation}
 Figure~\hyperlink{\ref{fig:MX_SHA}}{\ref{fig:MX_SHA}} displays the temperature dependence of this quantity at several representative electron fillings. While $\theta_{xy}^{s_z}$ is nearly temperature independent for generic fillings, as represented by the data at $n_{\mathrm{e}}=0.5$ and $1.5$, its magnitude is significantly enhanced at low temperatures for $n_{\mathrm{e}}=1$, where a pronounced peak is observed in Fig.~\hyperlink{fig:Spincond}{\ref{fig:Spincond}(c)}. The spin Hall angle reaches approximately $2.8h/(t\tau)$ at zero temperature. Assuming $t=0.1~{\mathrm{eV}}$ and $\tau=0.1~{\mathrm{ps}}$ as represetative values for chiral magnets \cite{Ishizuka_MnSi}, the spin Hall angle is estimated to exceed $110$~\% at zero temperature. It decreases with increasing temperature but remains approximately $34$~\% even near room temperature $\sim290$~K ($k_{\mathrm{B}}T/t=0.25$). These values are much larger than typical spin Hall angles reported for heavy metals \cite{SHE_Review}, indicating potentially efficient spin current generation via the MX. Note, however, that our estimates are based on an ideal model without accounting for specific scattering mechanisms. To obtain more realistic predictions, further investigations employing more sophisticated models beyond this approximate framework \cite{VertexInoue,VertexMishchenko,VertexChalaev,VertexDim} are necessary.

\begin{table*}[htbp]
\caption{Magnetization direction, spin space group, magnetic space group, and allowed spin current components for each spin texture. The bottom two rows list the symmetry-allowed spin polarization directions that give nonzero spin conductivities. Components in parentheses are allowed only when the SOC is present.}
\label{tab:spin_conductivity_summary}
\begin{ruledtabular}
\begin{tabular}{llccc}
 & &SkX & BmX & MX  \\
\hline
\multicolumn{2}{l}{Magnetization} & $z$ & $x$ &zero \\
\hline
\multicolumn{2}{l}{Spin space group} & $P^{4^1_{001}}4/^{1}m^{m_{100}}m^{m_{110}}m$ & $P^{4^1_{100}}4/^{1}m^{m_{001}}m^{m_{0\bar{1}1}}m$ & $P^{4^1_{001}}4/^1m^{m_{010}}m^{m_{\bar{1}10}}m(2_{100},2_{010},1)$\\
\hline
\multicolumn{2}{l}{Magnetic space group} & $P4m'm'$ & $Pm'$ & $P4b'm'$ \\
\hline
\multirow{2}{*}{Transverse}&intrinsic & $s_z$ & $s_x$, $(s_z)$ & $(s_z)$ \\
& dissipative & N/A & $(s_y)$ &  N/A  \\
\hline
\multirow{2}{*}{Longitudinal}& intrinsic & N/A & $(s_y)$ &  N/A  \\
 & dissipative & $s_z$ & $s_x$, $(s_z)$ & $(s_z)$ \\
\end{tabular}
\end{ruledtabular}
\end{table*}

\subsection{Symmetry analysis}
\label{subsec: symmetry}
 We now analyze the results presented in Secs.~\ref{subsec: Bands} and~\ref{subsec: transport} from the symmetry perspective. The symmetry of each spin texture and the corresponding allowed nonzero components of spin conductivity tensors are summarized in Table~\ref{tab:spin_conductivity_summary}. The symmetry of each spin texture is described by both the spin space group and the magnetic space group. The spin space group, where spin and spatial rotations are decoupled, provides a complete description of magnetic symmetry in the absence of SOC \cite{SSG1,SSG2,SSG3,SSG4,SSG5,SSG6,SSG7}. In contrast, the magnetic space group, where spin and spatial rotations are coupled, becomes necessary for a complete description when SOC is present. We determine the spin space group using \textsc{spinspg} library \cite{spinspg} and the magnetic space group using \textsc{spglib} library \cite{spglib1,spglib2} for each spin texture. We note that the symmetry-allowed nonzero components of spin conductivity tensors in the presence of SOC are consistent with the previous analysis based on the magnetic Laue group~\cite{MagLaue}.

 We begin with the SkX. The SkX belongs to the spin space group $P^{4^1_{001}}4/^{1}m^{m_{100}}m^{m_{110}}m$ and the magnetic space group $P4m'm'$. In this case, the magnetic space group alone is sufficient to explain the symmetries of spin-projected band structures, spin Berry curvatures, and spin conductivity tensors, regardless of whether the SOC is present. The magnetic space group $P4m'm'$ contains a fourfold rotation $4_{001}$ and combined operations of time-reversal and mirror reflection, $\mathcal{T}m_{100}$ and $\mathcal{T}m_{010}$, where $n_{\alpha}$ denotes an $n$fold rotation about the $[\alpha]$ direction ($\alpha$: Miller-indices), $\mathcal{T}$ is time-reversal operation, and $m_{\alpha}$ represents a mirror reflection with the $(\alpha)$ plane. Indeed, the band structure respects the fourfold rotational symmetry, and, as discussed in Secs.~\ref{subsec: Bands} and~\ref{subsec: transport}, the in-plane spin components of spin expectation value and spin Berry curvature exhibit symmetry with respect to $k_y=0$ and $k_x=0$, which are enforced by the symmetry with respect to $\mathcal{T}m_{100}$ and $\mathcal{T}m_{010}$, respectively. The symmetry with respect to $\mathcal{T}m_{100}$ imposes the relations
\begin{align}
    \hypertarget{eq:Tm100_const}{}
    \begin{split}
     \varepsilon_{n(k_x,k_y)}&=\varepsilon_{\tilde{n}(k_x,-k_y)},\\
     s_{x}(k_x,k_y;n)&=-s_{x}(k_x,-k_y;\tilde{n}),\\
     s_{y}(k_x,k_y;n)&=s_{y}(k_x,-k_y;\tilde{n}),\\
     s_{z}(k_x,k_y;n)&=s_{z}(k_x,-k_y;\tilde{n}),\\
     B^{s_x}_{xy}(k_x,k_y;n)&=-B^{s_x}_{xy}(k_x,-k_y;\tilde{n}),\\
     B^{s_y}_{xy}(k_x,k_y;n)&=B^{s_y}_{xy}(k_x,-k_y;\tilde{n}),\\
     B^{s_z}_{xy}(k_x,k_y;n)&=B^{s_z}_{xy}(k_x,-k_y;\tilde{n}),
    \end{split}
    \label{eq:Tm100_const}
\end{align}
 where $\tilde{n}$ denotes the band index of the partner connected to $n$ by the symmetry operation.
 From these relations, the spin conductivity tensor must satisfy
\begin{align}
    \hypertarget{eq:Tm100_const_spincond}{}
    \begin{split}
     s_x:~&\sigma^{s_x}_{xy}= \sigma^{s_x}_{yx}=\sigma^{s_x(\tau)}_{xx}= \sigma^{s_x(\tau)}_{yy}=0,\\
     s_y:~&\sigma^{s_y}_{xx}=\sigma^{s_y}_{yy}=\sigma^{s_y(\tau)}_{xy}=\sigma^{s_y(\tau)}_{yx}=0,\\
     s_z:~&\sigma^{s_z}_{xx}=\sigma^{s_z}_{yy}=\sigma^{s_z(\tau)}_{xy}=\sigma^{s_z(\tau)}_{yx}=0.
    \end{split}
    \label{eq:Tm100_const_spincond}
\end{align}
Similarly, the symmetry with respect to $\mathcal{T}m_{010}$ imposes 
\begin{align}
    \hypertarget{eq:Tm010_const}{}
    \begin{split}
     \varepsilon_{n(k_x,k_y)}&=\varepsilon_{\tilde{n}(-k_x,k_y)},\\
     s_{x}(k_x,k_y;n)&=s_{x}(-k_x,k_y;\tilde{n}),\\
     s_{y}(k_x,k_y;n)&=-s_{y}(-k_x,k_y;\tilde{n}),\\
     s_{z}(k_x,k_y;n)&=s_{z}(-k_x,k_y;\tilde{n}),\\
     B^{s_x}_{xy}(k_x,k_y;n)&=B^{s_x}_{xy}(-k_x,k_y;\tilde{n}),\\
     B^{s_y}_{xy}(k_x,k_y;n)&=-B^{s_y}_{xy}(-k_x,k_y;\tilde{n}),\\
     B^{s_z}_{xy}(k_x,k_y;n)&=B^{s_z}_{xy}(-k_x,k_y;\tilde{n})    
    \end{split}
    \label{eq:Tm010_const}
\end{align}
and therefore, the spin conductivity tensor satisfies
\begin{align}
    \hypertarget{eq:Tm010_const_spincond}{}
    \begin{split}
     s_x:~&\sigma^{s_x}_{xx}=\sigma^{s_x}_{yy}=\sigma^{s_x(\tau)}_{xy}=\sigma^{s_x(\tau)}_{yx}=0,\\
     s_y:~&\sigma^{s_y}_{xy}=\sigma^{s_y}_{yx}=\sigma^{s_y(\tau)}_{xx}=\sigma^{s_y(\tau)}_{yy}=0,\\
     s_z:~&\sigma^{s_z}_{xx}=\sigma^{s_z}_{yy}=\sigma^{s_z(\tau)}_{xy}=\sigma^{s_z(\tau)}_{yx}=0.
     \end{split}
     \label{eq:Tm010_const_spincond}
\end{align}
Taking into account the constraints of Eqs.~\hyperlink{eq:Tm100_const_spincond}{(\ref{eq:Tm100_const_spincond})} and \hyperlink{eq:Tm010_const_spincond}{(\ref{eq:Tm010_const_spincond})}, only the transverse intrinsic and longitudinal dissipative components with out-of-plane spin polarization are allowed to be nonzero for the SkX, as listed in Table~\ref{tab:spin_conductivity_summary}. Thus, the resulting spin conductivity tensor takes the form
\begin{align}
    \bm{\sigma}_{\mathrm{SkX}}^{s_z} =
\begin{pmatrix}
\sigma^{s_z(\tau)}_{xx} & \sigma^{s_z}_{xy} \\
-\sigma^{s_z}_{xy} & \sigma^{s_z(\tau)}_{xx}
\end{pmatrix}.
\end{align}
 This symmetry-based analysis is fully consistent with the numerical findings discussed in Sec.~\ref{subsec: transport}.

 Next, we consider the BmX. The BmX belongs to the spin space group $P^{4^1_{100}}4/^{1}m^{m_{001}}m^{m_{0\bar{1}1}}m$, which contains symmetry operations such as $\lbrace2_{100}||\mathcal{I}\rbrace$, $\lbrace4_{100}||4_{001}\rbrace$, and $\lbrace m_{001}||m_{100}\rbrace$, where $\mathcal{I}$ denotes spatial inversion, and the operations on the left and right of the double bar act in spin and real space, respectively. These symmetries ensure that, in the absence of SOC, only the same spin conductivity components as for SkX,  with spin polarization along the $x$ axis, are allowed to be nonzero.
 
 The same conclusion can be drawn more intuitively by considering the relation between the magnetic structures of SkX and BmX. In the absence of SOC, the Hamiltonians for SkX and BmX become identical once the 90\textdegree\ spin rotation that relates their magnetic structures is applied to the itinerant electrons. Therefore, as indicated in Sec.~\ref{subsec: transport}, the electronic structure and spin transport properties for the BmX coincide with those for the SkX with the spins rotated by 90\textdegree. As a result, only the spin conductivity components with spin polarization along the $x$ axis are allowed to be nonzero, consistent with the analysis based on the spin space group above. 
 
 In the presence of SOC, the symmetry of the system with BmX is fully described by the magnetic space group $Pm'$. This is much lower symmetry than that for the SkX, containing only two symmetry operations: the identity and $\mathcal{T}m_{010}$. Consequently, the symmetry constraints in Eq.~\hyperlink{eq:Tm100_const}{(\ref{eq:Tm100_const})} no longer apply to the BmX. This reduced symmetry is consistent with the observations that the spin-projected bands and spin Berry curvatures do not exhibit symmetric behavior with respect to $k_y=0$, as discussed in Secs.~\ref{subsec: Bands} and \ref{subsec: transport}. Regarding the spin conductivity tensor, the only remaining constraints are those given by Eq.~\hyperlink{eq:Tm010_const_spincond}{(\ref{eq:Tm010_const_spincond})}, allowing components that are forbidden in Eq.~\hyperlink{eq:Tm100_const_spincond}{(\ref{eq:Tm100_const_spincond})} to become nonzero, as listed in Table~\ref{tab:spin_conductivity_summary}. This is again fully consistent with the various nonzero components observed in Figs.~\hyperlink{fig:Spincond}{\ref{fig:Spincond}(b)} and \hyperlink{fig:Spincond}{\ref{fig:Spincond}(e)}. We note that, for other generic helicities, a further reduction of magnetic symmetry leads to additional nonzero components of the spin conductivity tensor; see the Appendix~\ref{subsec:appendix_BmX} for details.

 Finally, we turn to the MX. As in the BmX case, the symmetry of MX changes drastically depending on the presence or absence of SOC, resulting in significant differences in its physical properties. In the absence of SOC, the symmetry of MX is described by the spin space group $P^{4^1_{001}}4/^1m^{m_{010}}m^{m_{\bar{1}10}}m(2_{100},2_{010},1)$, possessing symmetry operations $\mathcal{O}_1=\lbrace2_{001}||m_{001}\rbrace$ and $\mathcal{O}_2=\lbrace2_{100}||m_{001}|\frac{1}{2},0\rbrace$,
 where the translation components appear to the right of the single bar. As discussed in Ref.~\cite{Zhentao-MX}, the symmetries with respect to $\mathcal{O}_1$ and $\mathcal{O}_2$ protect the twofold spin degeneracy of electronic bands throughout the Brillouin zone. In addition, these symmetries forbid all components of the spin conductivity tensor, resulting in vanishing spin conductivity. These are consistent with the numerical results shown in Secs.~\ref{subsec: Bands} and \ref{subsec: transport}. 
 
 In contrast, when the SOC is switched on, the system with MX is described by the magnetic space group $P4b'm'$. Note that we take into account the breaking of the mirror symmetry with respect to the $xy$ plane due to the out-of-plane potential gradient, which causes Rashba SOC. The corresponding magnetic point group is $4m'm'$, which is identical to that of the SkX. Therefore, the symmetry of the spin-projected band structure and the spin Berry curvature is the same as that for the SkX, leading to an identical tensor form of the spin conductivity. The key difference from the SkX, however, lies in the presence of nonsymmorphic symmetries in the MX, specifically glide operations combined with time-reversal operation: $\mathcal{O}_3=\lbrace\mathcal{T}m_{010}|\frac{1}{2},0\rbrace$ and $\mathcal{O}_4=\lbrace\mathcal{T}m_{100}|0,\frac{1}{2}\rbrace$. As discussed in Ref.~\cite{Zhentao-MX}, these nonsymmorphic symmetries protect spin degeneracy along the Brillouin zone boundary, with $\mathcal{O}_3$ and $\mathcal{O}_4$ responsible for the degeneracy on the XM and YM lines, respectively. This symmetry-protected band degeneracy thus offers a consistent explanation for the divergent spin Berry curvature $B^{s_z}_{xy}$ around the XM line, which underlies the large spin Hall conductivity discussed in Sec.~\ref{subsec: transport}. For other generic helicities, as such nonsymmorphic symmetries do not exist, the spin degeneracy is not protected, and thus the spin Hall conductivity is suppressed as discussed in the Appendix \ref{subsec:appendix_MX}.
\section{Summary and perspectives}
\label{sec: Summary}
 In summary, we have studied spin current generation in a two-dimensional model in which itinerant electrons are coupled to a topological spin texture, using linear response theory with particular attention to the effect of SOC. We considered three distinct textures that share a common out-of-plane emergent magnetic field but differ in their magnetizations: SkX with out-of-plane magnetization, BmX with in-plane magnetization, and MX with zero net magnetization. We clarified that SOC significantly influences the electronic and transport properties for the BmX and MX, while the SkX exhibits qualitatively similar behavior regardless of the presence of SOC. In the absence of SOC, the BmX shares the same band dispersion as the SkX, and the MX shows no spin splittings. When the SOC is introduced, the band structure for the BmX exhibits lower symmetry compared to the SkX, and the MX develops momentum-dependent spin splittings while maintaining degeneracy on the Brillouin zone boundary. These differences in the band structure lead to distinct spin transport properties. For the BmX, although only the spin conductivity components polarized along the magnetization direction, as for the SkX, are nonzero without SOC, multiple components become nonzero in the presence of SOC. In contrast, the MX, which generates no spin current without SOC, produces sizable spin currents with out-of-plane polarization at a certain filling once SOC is introduced, despite its zero net magnetization. Its large spin-polarized current originates from an enhanced spin Berry curvature, which arises from spin degeneracy along the Brillouin zone boundary. We have also demonstrated that these results can be understood qualitatively in terms of magnetic symmetry through analysis based on the spin space group and magnetic space group.

 Our results may have implications for experimental realizations and potential applications in spintronics devices. The BmX, for instance, has been experimentally realized in a thin film grown on a substrate through the engineered Dzyaloshinskii--Moriya interaction \cite{BmXexp}, suggesting that the spin current generation mechanism discussed in this work could be accessible in experiments. From an application viewpoint, the BmX is particularly promising, as it can generate spin current in both longitudinal and transverse directions with nonzero spin polarizations across all components. This richness offers the flexibility to control magnetic memories with arbitrary spin directions. In contrast, the MX considered in this study, which has no net magnetization, has not yet been observed experimentally, to the best of our knowledge. Nevertheless, theoretical models hosting the MX have been reported \cite{Zhentao-MX,Hayami-MX}, suggesting that their experimental realization may be within reach. For applications, the MX offers a unique advantage: it can generate sizable spin-polarized currents even without producing stray magnetic fields, making it an attractive candidate for integration into densely packed spintronic device architectures. Our findings broaden the scope of spintronics without net magnetization, which has recently been explored in systems such as altermagnets \cite{almag1,almag_Libor1,almag_Libor2}.

A promising theoretical direction for future research is to explore the impact of dynamical spin excitations on spin transport. In our analysis, the topological spin textures are treated as static backgrounds; however, dynamical excitations may induce more intriguing responses. For instance, spin dynamics can generate emergent electric fields, which directly affect transport properties of itinerant electrons \cite{EMF1,EMF2,EMF3,SMF1}. In addition, topological spin textures may host characteristic spin excitations, exemplified as topological magnons with nonzero Chern numbers \cite{Topomagnon1,Topomagnon2,Topomagnon3}. These dynamical effects may lead to novel spin transport phenomena unique to topological magnets. It is also important to investigate the temperature and magnetic field dependencies of spin current generation by explicitly incorporating the stability mechanisms of the topological spin textures, which were not considered in this study. In particular, analysis near phase transitions will be crucial, as the phase competition and associated fluctuations can significantly modify or even enhance spin current responses. Another promising extension is to broaden the scope to a wider class of topological spin textures beyond those investigated in this study, including three-dimensional ones such as magnetic hedgehogs and hopfions. Although a previous study reported a vanishing spin Hall conductivity for hopfions \cite{OHE_hop}, our results suggest that incorporating SOC may qualitatively modify their spin transport properties. Such investigations may further expand the possibilities of spin current generation via topological magnetism.

\section*{Acknowledgment}
A. K. thanks K. Fukui and S. Ikegami for valuable discussions. This work was supported by the JPSJ KAKENHI (Grants No. JP25H01247, JP22K13998, JP23K25816) and JST PRESTO (No. JPMJPR2595). A. K. was supported by the Program for Leading Graduate Schools (MERIT-WINGS).

\appendix
\begin{figure*}[t]
    \centering
    \hypertarget{fig:Heli_textures}{}
    \includegraphics[width=1\textwidth]{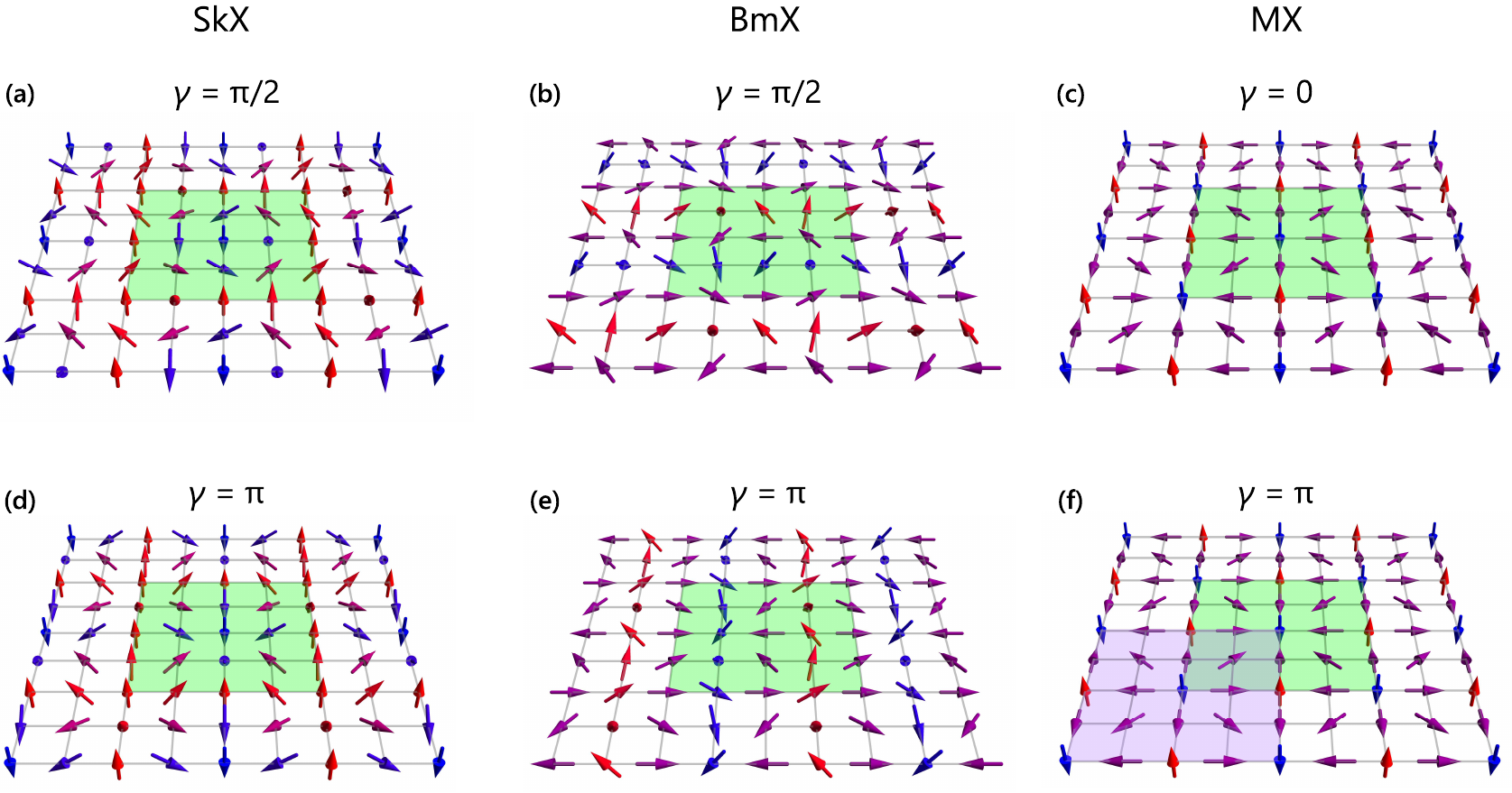}
    \caption{Spin configurations of topological spin textures with representative helicities not considered in the main text. (a), (b) Bloch-type SkX and BmX with $\gamma=\pi/2$, respectively. (c) N\'{e}el-type MX with $\gamma=0$. (d), (e) N\'{e}el-type SkX and BmX with $\gamma=\pi$, respectively. (f) N\'{e}el-type MX with $\gamma=\pi$. The green-shaded squares represent the magnetic unit cell, which consists of $4 \times 4$ sites. In panel (f), the purple-shaded square also denotes a magnetic unit cell, where the spin configuration is identical to that in the green-shaded cell in panel (c).}
    \label{fig:Heli_textures}
\end{figure*}

\section{Results for other helicities}
\label{apx:Results for other helicities}
We discussed the N\'{e}el-type SkX and BmX with $\gamma=0$, and the Bloch-type MX with $\gamma=\pi/2$ in the main text. In this appendix, we investigate spin conductivities for other helicities $\gamma$, exemplified by the textures in Fig.~\ref{fig:Heli_textures}.

 In the absence of SOC, the spin conductivity for each texture is independent of the helicity $\gamma$. This can be understood most straightforwardly by considering the SkX case. For a nonzero helicity $\gamma$, the corresponding Hamiltonian can be obtained from that for $\gamma=0$ by applying a global spin rotation about the $z$ axis by an angle $\gamma$. Such a spin rotation does not affect the $z$ component of spin polarization, thereby preserving the $s_z$ components of spin conductivity. Since the $s_x$ and $s_y$ components vanish for $\gamma=0$, they remain zero after the spin rotation. Consequently, the spin conductivity for the SkX exhibits no dependence on $\gamma$ in the absence of SOC.
 The same argument applies to the BmX and MX, leading to the same independence of $\gamma$. Therefore, in the following, we consider each spin texture in the presence of SOC, with its strength set to $\lambda_{\mathrm{R}}=0.1t$.

\begin{figure*}[h]
    \centering
    \hypertarget{fig:Heli_all}{}
    \includegraphics[width=\linewidth]{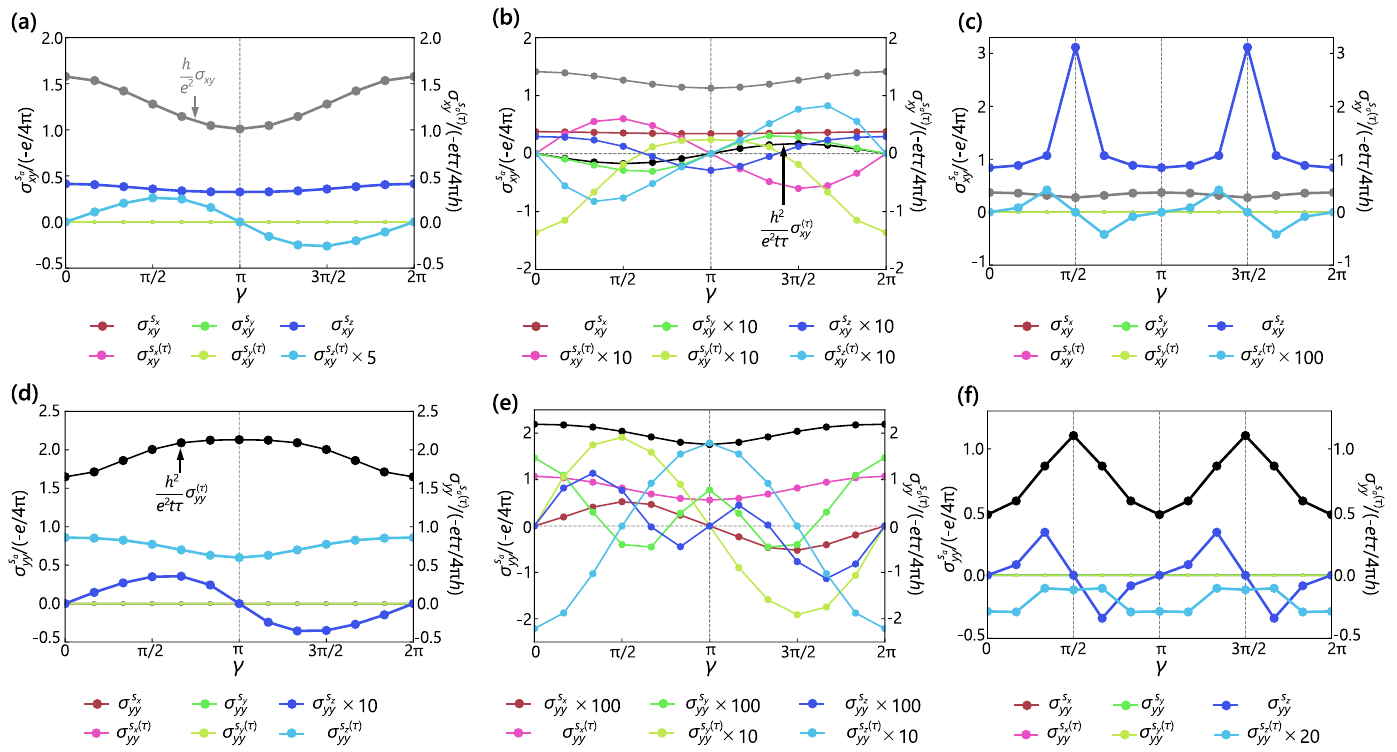}
    \caption{Helicity dependencies of the charge and spin conductivities for three spin textures in the presence of SOC. The left, center, and right columns correspond to SkX, BmX, and MX, respectively: (a)--(c) charge Hall conductivities $\sigma_{xy}$ (intrinsic, plotted as the dimensionless quantity $h\sigma_{xy}/e^2$) and $\sigma^{(\tau)}_{xy}$ (dissipative, plotted as the dimensionless quantity $h^2\sigma^{(\tau)}_{xy}/e^2t\tau$), and spin Hall conductivities $\sigma^{s_a}_{xy}$ (intrinsic, left axes) and $\sigma^{s_a(\tau)}_{xy}$ (dissipative, right axes), and (d)--(f) longitudinal charge conductivity $\sigma^{(\tau)}_{yy}$ (dissipative, plotted as the dimensionless quantity $h^2\sigma^{(\tau)}_{yy}/e^2t\tau$), and spin conductivities $\sigma^{s_a}_{yy}$ (intrinsic, left axes) and $\sigma^{s_a(\tau)}_{yy}$ (dissipative, right axes). In each panel, the intrinsic and dissipative contributions to the charge conductivity are shown in gray and black, respectively, while the spin conductivities are plotted in different colors according to the spin polarization direction $a=x,y,z$ and the type of contribution (intrinsic or dissipative). The legends for the spin conductivities are shown below each panel. For better visibility, some components of spin conductivity are appropriately rescaled, as indicated in the legends.}
    \label{fig:Heli_all}
\end{figure*}
\subsection{SkX}

We now examine the SkX in the presence of SOC. Figures~\hyperlink{fig:Heli_all}{\ref{fig:Heli_all}(a)} and \hyperlink{fig:Heli_all}{\ref{fig:Heli_all}(d)} show the helicity dependencies of charge and spin conductivities at a fixed electron filling $n_{\mathrm{e}}=2.5$. The SkX, for any $\gamma$, the only nonzero components of spin conductivity are those with out-of-plane spin polarization. For the N\'{e}el-type SkX with $\gamma=\pi$, both intrinsic $\sigma^{s_z}_{xy}$ and dissipative  $\sigma^{s_z(\tau)}_{yy}$ are nonzero, as for $\gamma=0$ discussed in Sec.~\ref{subsec: transport}, while their magnitudes are different. By contrast, for other helicities, intrinsic $\sigma^{s_z}_{yy}$ and dissipative $\sigma^{s_z(\tau)}_{xy}$ also become nonzero. These additional components arise from a reduction in magnetic symmetry. Once $\gamma$ deviates from $0$ or $\pi$, the magnetic space group is lowered from $P4m'm'$ to $P4$, allowing a more general form of the spin conductivity tensor as
 \begin{align}
    \bm{\sigma}_{\mathrm{SkX}}^{s_z} =
\begin{pmatrix}
\sigma^{s_z}_{xx}+\sigma^{s_z(\tau)}_{xx} & \sigma^{s_z}_{xy}+\sigma^{s_z(\tau)}_{xy} \\
-\sigma^{s_z}_{xy}-\sigma^{s_z(\tau)}_{xy} & \sigma^{s_z}_{xx}+\sigma^{s_z(\tau)}_{xx}
\end{pmatrix},
\end{align}
 which is fully consistent with the results shown in Figs.~\hyperlink{fig:Heli_all}{\ref{fig:Heli_all}(a)} and \hyperlink{fig:Heli_all}{\ref{fig:Heli_all}(d)}. In addition, both charge and spin conductivities exhibit symmetry with respect to $\gamma=\pi$: $\sigma_{xy}$, $\sigma^{s_z}_{xy}$, $\sigma^{(\tau)}_{yy}$ and $\sigma^{s_z(\tau)}_{yy}$ are even functions about $\gamma=\pi$, while $\sigma^{(\tau)}_{xy}$ and $\sigma^{s_z}_{yy}$ are odd. This behavior can be understood from the fact that the SkX with $\gamma$ and $2\pi-\gamma$ are related by the symmetry operation $\mathcal{T}m_{010}$.

\subsection{BmX}
\label{subsec:appendix_BmX}
We next discuss the results for the BmX, shown in Figs. ~\hyperlink{fig:Heli_all}{\ref{fig:Heli_all}(b)} and \hyperlink{fig:Heli_all}{\ref{fig:Heli_all}(e)}. Similar to the SkX, the N\'{e}el-type BmX with $\gamma=\pi$ exhibits the same set of nonzero spin conductivity components as the case with $\gamma=0$, although their magnitudes differ. Away from these N\'{e}el-type helicities, all components of the spin conductivity tensor become nonzero, and the dissipative charge Hall conductivity $\sigma^{(\tau)}_{xy}$ also appears. These differences also stem from a reduction in magnetic symmetry. For $\gamma \neq 0$ (or equivalently $2\pi$) and $\pi$, the magnetic space group is lowered to $P1$, which does not impose any constraint on either the charge or spin conductivities. Consequently, all components are allowed to be nonzero. In addition, as for the SkX, both charge and spin conductivities exhibit symmetry with respect to $\gamma=\pi$ in Figs.~\hyperlink{fig:Heli_all}{\ref{fig:Heli_all}(b)} and \hyperlink{fig:Heli_all}{\ref{fig:Heli_all}(e)}, as the BmX for $\gamma$ and $2\pi-\gamma$ are related by the symmetry operation $\mathcal{T}m_{010}$. We note that, if the potential gradient perpendicular to the $xy$ plane that induces Rashba SOC were absent, the magnetic space group of the Bloch-type BmX with $\gamma=\pi/2$ and $3\pi/2$ would be $P2'$, distinguishing these special helicities from other generic cases discussed above.

\subsection{MX}
\label{subsec:appendix_MX}

\begin{figure*}[htbp]
    \centering
    \hypertarget{fig:HeliMX_bandberry}{}
    \includegraphics[width=0.8\linewidth]{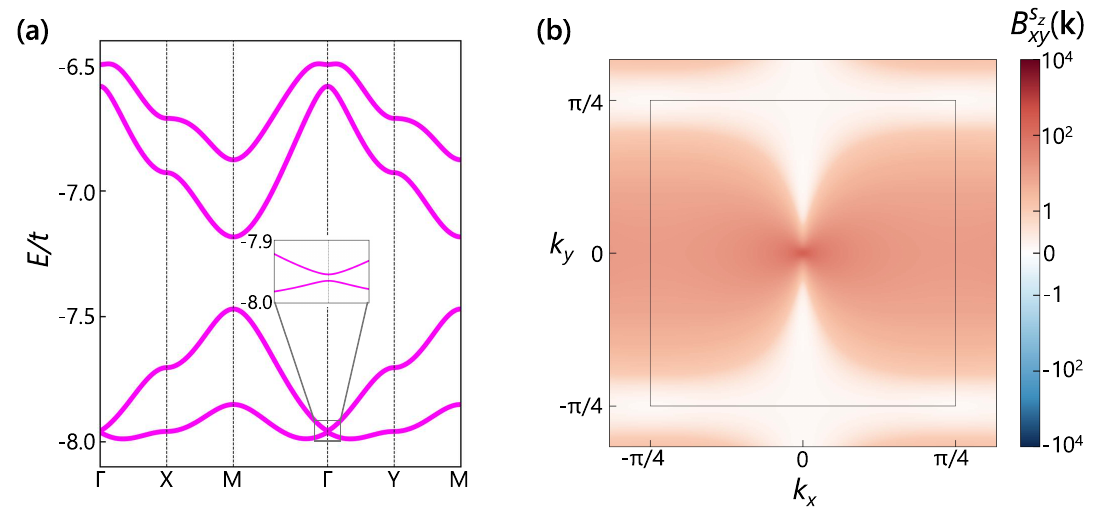}
    \caption{Electronic band structure and spin Berry curvature for the N\'{e}el-type MX with $\gamma=0$ in the presence of SOC. (a) The four lowest-energy bands on the high-symmetry lines. The inset shows an enlarged view around the $\Gamma$ point. (b) The spin Berry curvature $B^{s_z}_{xy}$ for the lowest-energy band. The black square indicates the magnetic Brillouin zone boundary.}
    \label{fig:HeliMX_bandberry}
\end{figure*}

We next turn to the MX. Unlike the SkX and BmX, the magnetic structure of MX has a helicity period of $\pi$. For instance, the spin configuration with $\gamma=0$ in the green-shaded magnetic unit cell in Fig.~\hyperlink{fig:Heli_textures}{\ref{fig:Heli_textures}(c)} is identical to that with $\gamma=\pi$ in the purple-shaded magnetic unit cell in Fig.~\hyperlink{fig:Heli_textures}{\ref{fig:Heli_textures}(f)}. Similarly, for any $\gamma$, the spin configurations with $\gamma$ and $\gamma+\pi$ are related by a lattice translation and are therefore regarded as equivalent textures. 

Figures~\hyperlink{fig:Heli_all}{\ref{fig:Heli_all}(c)} and \hyperlink{fig:Heli_all}{\ref{fig:Heli_all}(f)} show the helicity dependencies of the charge and spin conductivities at $n_{\mathrm{e}}=1$, where the intrinsic spin Hall conductivity $\sigma^{s_z}_{xy}$ is enhanced for $\gamma=\pi/2$ as indicated in Sec.~\ref{subsec: transport}. For $\gamma=0$ (or equivalently $\gamma=\pi$ and $2\pi$) and $\gamma=\pi/2$ (or $\gamma=3\pi/2$), the nonzero components of spin conductivity are the intrinsic $\sigma^{s_z}_{xy}$ and dissipative $\sigma^{s_z(\tau)}_{yy}$, whereas additional components, $\sigma^{s_z(\tau)}_{xy}$ and $\sigma^{s_z}_{yy}$, appear for other $\gamma$. This difference is also explained by the difference in magnetic symmetry, as for the SkX and BmX discussed above. Specifically, the magnetic space group is $P4m'm'$ for $\gamma=0$ and $P4b'm'$ for $\gamma=\pi/2$, both belonging to the same magnetic point group $4m'm'$, and therefore allowing the same set of components. In contrast, for other $\gamma$, the symmetry is lowered to $P4$, leading to the appearance of additional components. In addition, both charge and spin conductivities exhibit symmetry with respect to $\gamma=\pi/2$, as the spin configurations with $\gamma$ and $\pi-\gamma$ related by the symmetry operation $\lbrace\mathcal{T}m_{100}|0,\frac{1}{2}\rbrace$, as shown in Figs.~\hyperlink{fig:Heli_all}{\ref{fig:Heli_all}(c)} and \hyperlink{fig:Heli_all}{\ref{fig:Heli_all}(f)}.

Although the intrinsic spin Hall conductivity $\sigma^{s_z}_{xy}$ remains nonzero across all helicities, its magnitude takes the maximum value at $\gamma=\pi/2$ (or $3\pi/2$) and is suppressed for other helicities. This is because, for generic helicities other than $\gamma=\pi/2$, the nonsymmorphic symmetry that preserves spin degeneracy at the Brillouin zone boundary is absent. Therefore the enhancement of the spin Berry curvature discussed in Secs.~\ref{subsec: transport} and \ref{subsec: symmetry} does not occur. For example, at $\gamma=0$, the electronic band structure exhibits no spin degeneracy as shown in Fig.~\hyperlink{fig:HeliMX_bandberry}{\ref{fig:HeliMX_bandberry}(a)}, and thus the spin Berry curvature takes a modest value near the $\Gamma$ point where the two lowest-energy bands approach each other [see the inset in Fig.~\hyperlink{fig:HeliMX_bandberry}{\ref{fig:HeliMX_bandberry}(a)}], without any pronounced enhancement [Fig.~\hyperlink{fig:HeliMX_bandberry}{\ref{fig:HeliMX_bandberry}(b)}]. This leads to the reduced intrinsic spin Hall conductivity $\sigma^{s_z}_{xy}$ compared to that for $\gamma=\pi/2$.

\clearpage
\bibliography{refs}
\end{document}